\documentclass[fleqn,usenatbib]{mnras}

\usepackage{newtxtext,newtxmath}

\usepackage[T1]{fontenc}
\usepackage{ae,aecompl}

\usepackage{graphicx}	
\usepackage{amsmath}	

\usepackage{amssymb}	
\usepackage{verbatim}
\usepackage{multirow}
\usepackage{threeparttable}
\usepackage{relsize}
\usepackage{color}

\usepackage{caption}
\usepackage{subcaption}

\usepackage[splitrule]{footmisc} 
\def\gtra{\mathrel{\hbox{\rlap{\hbox{\lower4pt\hbox{$\sim$}}}\hbox{$>$}}}}
\def\lessa{\mathrel{\hbox{\rlap{\hbox{\lower4pt\hbox{$\sim$}}}\hbox{$<$}}}}

\newcommand{\angstrom}{\textup{\AA}}

\newcommand{\fesclya}{f_{\rm esc, Ly\alpha}}
\newcommand{\fesclyc}{f_{\rm esc, LyC}}
\newcommand{\vsep}{v_{\rm sep}}
\newcommand{\fcen}{f_{\rm cen}}
\newcommand{\bLAH}{b_{\rm LAH}}
\newcommand{\sred}{\sigma_{\rm red}}
\newcommand{\NHI}{N_{\rm H~\textsc{i}}}

\usepackage[normalem]{ulem}

\definecolor{darkgreen}{rgb}{0.13, 0.55, 0.13}
\definecolor{orange}{rgb}{0.8, 0.15, 0.13}

\newcommand{\aref}[1]{\hyperref[#1]{Appendix~\ref{#1}}}

\usepackage{scalerel,tikz}
\usetikzlibrary{svg.path}
\definecolor{orcidlogocol}{HTML}{A6CE39}
\tikzset{orcidlogo/.pic={
 \fill[orcidlogocol] svg{M256,128c0,70.7-57.3,128-128,128C57.3,256,0,198.7,0,128C0,57.3,57.3,0,128,0C198.7,0,256,57.3,256,128z};
 \fill[white] svg{M86.3,186.2H70.9V79.1h15.4v48.4V186.2z}
 svg{M108.9,79.1h41.6c39.6,0,57,28.3,57,53.6c0,27.5-21.5,53.6-56.8,53.6h-41.8V79.1z M124.3,172.4h24.5c34.9,0,42.9-26.5,42.9-39.7c0-21.5-13.7-39.7-43.7-39.7h-23.7V172.4z}
 svg{M88.7,56.8c0,5.5-4.5,10.1-10.1,10.1c-5.6,0-10.1-4.6-10.1-10.1c0-5.6,4.5-10.1,10.1-10.1C84.2,46.7,88.7,51.3,88.7,56.8z};
}}
\newcommand\orcidicon[1]{\href{https://orcid.org/#1}{\mbox{\scalerel*{
\begin{tikzpicture}[yscale=-1,transform shape]
\pic{orcidlogo};
\end{tikzpicture}
}{|}}}}

\title[Ly$\alpha$ as a probe of LyC escape]{Ly$\alpha$ emission as a sensitive probe of feedback-regulated LyC escape from dwarf galaxies}

\author[Yuan et al.]{Yuxuan Yuan$^{1}$~\orcidicon{0000-0001-6816-0682}\thanks{E-mail: yy503@cam.ac.uk (YY)}, Sergio Martin-Alvarez$^{2}$~\orcidicon{0000-0002-4059-9850}, Martin G. Haehnelt$^{1}$, Thibault Garel$^{3}$, \newauthor and Debora Sijacki$^{1}$
\\
$^{1}$ Institute of Astronomy and Kavli Institute for Cosmology, University of Cambridge, Madingley Road, Cambridge CB3 0HA, UK\\
$^{2}$ Kavli Institute for Particle Astrophysics \& Cosmology (KIPAC), Stanford University, Stanford, CA 94305, USA\\
$^{3}$ Observatoire de Genève, Université de Genève, Chemin Pegasi 51, 1290 Versoix, Switzerland
}

\date{MNRAS, submitted}

\pubyear{2023}

\begin{document}
\label{firstpage}
\pagerange{\pageref{firstpage}--\pageref{lastpage}}
\maketitle

\begin{abstract}
Ly$\alpha$ emission is an exceptionally informative tracer of the life cycle of evolving galaxies and the escape of ionising photons. However, theoretical studies of Ly$\alpha$ emission are often limited by insufficient numerical resolution, incomplete sets of physical models, and poor line-of-sight (LOS) statistics. To overcome such limitations, we utilize here the  novel \textit{PANDORA} suite of high-resolution dwarf galaxy simulations that include a comprehensive set of state-of-the-art physical models for ionizing radiation, magnetic fields, supernova feedback and cosmic rays. We post-process the simulations with the radiative transfer code \textsc{RASCAS} to generate synthetic observations and compare to observed properties of Ly$\alpha$ emitters. Our simulated  Ly$\alpha$ haloes are more extended than the spatial region from which the intrinsic emission emanates and our spatially resolved maps of spectral parameters of the Ly$\alpha$ emission are very sensitive to the underlying spatial distribution and kinematics of neutral hydrogen. Ly$\alpha$ and LyC emission display strongly varying signatures along different LOS depending on how each LOS intersects low-density channels generated by stellar feedback. Comparing galaxies simulated with different physics, we find the Ly$\alpha$ signatures to exhibit systematic offsets determined by the different levels of feedback strength and the clumpiness of the neutral gas. Despite this variance, and regardless of the different physics included in each model, we find universal correlations between Ly$\alpha$ observables and LyC escape fraction, demonstrating a robust connection between Ly$\alpha$ and LyC emission. Ly$\alpha$ observations from a large sample of dwarf galaxies should thus give strong constraints on their stellar feedback-regulated LyC escape and confirm their important role for the reionization of the Universe.
\end{abstract}


\section{Introduction}
\label{intro}

During its first billion years, the Universe transitioned from neutral hydrogen to ionized hydrogen in the Epoch of Reionization (EoR). This phase transition was driven by the ionizing photons (photons with wavelength $\lambda<912 \angstrom$, also called Lyman continuum or LyC) emitted from the first galaxies and quasars \citep{Loeb&Barkana01, OShea15, Ocvirk16, Mutch16a, Stark16, Rosdahl18, Kulkarni19, Ma20}. However, understanding how this happens in detail and how large the contribution from active galactic nuclei (AGN) to hydrogen reionization remain open questions. Several previous studies have come to the consensus that, since the number of quasars declines rapidly with increasing redshift, photons from the first star-forming galaxies are the primary source of reionization  \citep{Madau99, Fontanot12, Qin17c, Faucher-Giguere20, Trebitsch21, Yeh23}, although recent studies by \citet{Grazian22, Grazian23, Harikane23, Maiolino23} have reinvigorated the hypothesis that AGN may contribute significantly to hydrogen reionization. 

The ionizing photon budget for the EoR is mainly determined by three factors: UV luminosity density $\rho_{\rm UV}$, conversion between UV and ionizing photon production $\xi_{\rm ion}$, and escape fraction $f_{\rm esc}$. Among these quantities, the escape fraction of ionizing photons from galaxies is probably the most uncertain. Direct measurement of LyC escape in observations at high redshift (z>6) is impossible, as the intervening intergalactic medium (IGM) becomes opaque. Observational strategies therefore have resorted to direct measurements of LyC escape of low redshift galaxies selected to be analogues of high-z galaxies. Several indirect methods have been suggested for estimating LyC escape based on a combination of other emission lines. These indirect methods are then applied to high-z galaxies with the assumption that the emission line physics is invariant across redshift. For example, \citet{Zackrisson13, Zackrisson17} predict the escape fraction based on H$\beta$ equivalent width (EW) and UV slope. Other indicators include high O32 \citep{Izotov16, Bassett19, Nakajima20, Katz20b, Barrow20}, and S \textsc{ii} deficit \citep{Wang19, Wang21}. \citet{Katz20b} employed machine learning techniques to perform an extensive emission line analysis on a cosmological zoom-in simulation and proposed a multi-line estimator including [O \textsc{iii}] and [C \textsc{ii}]. Resonant lines are another class of powerful predictive tools for this purpose, where the most promising one is Ly$\alpha$. Preliminary analysis has shown that Ly$\alpha$ emission can identify LyC leakers based on peak separation \citep{Verhamme15, Izotov18} and EWs \citep{Steidel18, Pahl23}. 

More than 50 years ago, \citet{Partridge&Peebles67} predicted that massive stars in high-z galaxies ionize the surrounding gas and that their subsequent recombination generates strong Ly$\alpha$ emission. Such emission was first detected as Ly$\alpha$ blobs (LAB) and Ly$\alpha$ nebulae around high-$z$ quasars \citep{McCarthy87, Steidel00}, with a large scale length of $\sim 10 -100$~kpc. Later surveys utilized narrow-band imaging to obtain ultra-deep, blind samples of stacked Ly$\alpha$ haloes around star-forming galaxies \citep{Hayashino04, Steidel11}. These haloes show more extended Ly$\alpha$ emission than the optical and UV counterparts of the galaxies. For this reason, Ly$\alpha$ emission has been shown to be a sensitive tracer of not only the interstellar medium (ISM), but also the circumgalactic medium (CGM) of high-z galaxy populations \citep{Rosdahl&Blaizot12, Byrohl21, Garel21, Mitchell21}. The discovery of individual extended Ly$\alpha$ haloes around bright quasi-stellar objects (QSO) by \citet{Cantalupo14}, suggests the presence of a neutral clumpy CGM. 

With the greater sensitivities, higher resolutions and larger covering areas of the new observational facilities, there are now more surveys than ever before providing us with access to the Ly$\alpha$ emission from star-forming galaxies across cosmic time \citep{Hayes15, Ouchi20}. In the local Universe, prominent examples are the Lyman Alpha Reference Sample (LARS) survey \citep{Hayes13} performed with the Hubble Space Telescope (HST), and the X-SHOOTER Ly$\alpha$ survey at $z=2$ \citep[XLS-z2,][]{Matthee21, Matthee22}  performed with the \textsc{Very Large Telescope} (VLT). At high $z \sim 3-7$, early work by \citet{Rauch08} identified a large sample of faint extended Ly$\alpha$ emitters and discussed their likely relation to damped Ly$\alpha$ systems (DLA) or high column density Lyman limit systems (LLS). About a decade later, the MUSE-WIDE \citep{Urrutia19} and MUSE-UltraDeep surveys \citep{Bacon17, Leclercq17} gave us access to much larger samples of faint Ly$\alpha$ haloes, exploiting the integral field unit (IFU) spectrographs on the Multi Unit Spectroscopic Explorer (MUSE) and its large field of view. At even higher redshift, Ly$\alpha$ observations suffer from IGM attenuation and Ly$\alpha$ emission becomes harder to detect. Nonetheless, the JADES survey \citep{Saxena24, Eisenstein23, Bunker23} and CEERS survey \citep{Jung24, Tang23} conducted by JWST has expanded observations at high redshift to fainter sources and probed diffuse Ly$\alpha$ emission from the environment of the galaxies.

The propagation of Ly$\alpha$ photons in the Universe is complex, as Ly$\alpha$ is a resonant line and photons will undergo resonant scattering in the surrounding neutral gas and absorption by dust \citep{Dijkstra19}. Recently, and thanks to the advent of more powerful radiative transfer tools, the number of theoretical studies investigating Ly$\alpha$ emission using numerical simulations has been increasing. Early analytical models provided physical insight into the relation between line profile, density and velocity structures, and photons escape assuming slab and spherical geometries \citep{Harrington73, Neufeld90, Loeb&Rybicki99, Hansen06, Dijkstra06}.  The next generation of models conducted were idealized simulations with configurations of expanding spherical shells  \citep{Verhamme08, Barnes10, Gronke15, Yang17}, clumpy multiphase-media \citep{Hansen06, Dijkstra12, Laursen13, Gronke17a} and outflow-driven cavities \citep{Behrens14}. These simulations dug deeper into the diversity of Ly$\alpha$ profiles, and the dependence of Ly$\alpha$ on various galaxy properties and redshift over a larger volume of parameter space. However, fully understanding the physical processes regulating Ly$\alpha$ emission requires self-consistent hydro-dynamical simulations \citep{Laursen09a, Faucher-Giguere10, Yajima12, Smith15, Aaron_Smith19, Gronke18, Mitchell21}. This is due to the complex distribution of the Ly$\alpha$ sources and the scattering medium, which are affected by multiple factors including star formation \citep{Kimm17}, stellar feedback \citep{Kimm&Cen14} and galaxy mergers \citep{Witten24}, as well as the properties of the ISM and CGM.

Over the past decade, we have seen significant improvement in the modelling of galaxy formation in the EoR. Advances in radiation hydrodynamics give us a better understanding of the role of inhomogeneous reionization driven by stars beyond the assumption of a spatially uniform ultraviolet background (UVB) \citep{Bolton17, Ma18a, Ma18b, Rosdahl18, Ma20, Katz20a, Ocvirk20, Kannan22a, Garaldi22, Puchwein23}. Improvement of the subgrid modelling in galaxy formation simulations, such as star formation \citep{Padoan&Nordlund11, Federrath12, Trebitsch17, Kimm17}, on-the-fly radiative transfer (RT), supernovae (SN) feedback \citep{Kimm&Cen14, CG_Kim15a, Martizzi15, Hopkins18a, Smith19} and cosmic ray (CR) feedback \citep{Salem&Bryan14, Jiang&Oh18, Dubois19, Hopkins21a, Hopkins21c}, combined with the increasing resolution that is now starting to spatially resolve the ISM, give us a better view of the multi-scale structures of galaxies and the constraining power of observations. For example, \citet{Laursen09a, Faucher-Giguere10, Yajima12, Smith15, Gronke18, Aaron_Smith19, Aaron_Smith22} predict that feedback processes and/or turbulence clear low density channels for the Ly$\alpha$ photons to escape from the ISM. 

Recently, it has become feasible to study the role of Ly$\alpha$ in a cosmological context. Both \citet{Byrohl21} and \citet{Mitchell21} show that resonant scattering and satellite emission results in extended Ly$\alpha$ haloes with flattened radial profiles consistent with results from the MUSE survey. \citet{Garel21} analysed Ly$\alpha$ emission for thousands of galaxies in the SPHINX simulation \citep{Rosdahl18, Rosdahl22, Katz23} to gain a statistical view of the LAEs at high $z$. \citet{Garel21} further showed that faint, low-mass galaxies are more likely to exhibit a blue peak than bright, massive ones, and that it is the IGM transmission that drives the apparent redshift evolution of LAEs. \citet{Maji22} conducted a follow-up study and proposed multivariate linear models relating the physical and Ly$\alpha$ properties of galaxies to LyC luminosities and escape. However, both \citet{Garel21} and \citet{Maji22} focused on  angle-averaged results and neglected LOS variations. \citet{Blaizot23} generated a large bank of realistic Ly$\alpha$ spectra along a number of LOS and snapshots between $z=4.18\to 3$ for a single simulated low-mass galaxy. They also took into account aperture effects. They then investigated the link between Ly$\alpha$ properties and gas cycles and found that these spectra can fit almost all the locally observed spectra in the Ly$\alpha$ spectral database (LASD) \citep{Runnholm21}. Note, however, that they did not relate Ly$\alpha$ observables to the escape of ionising photons. Furthermore, all above-mentioned simulations only considered only one specific physical model and did not explore the combined effects of RHD with SN feedback, magnetic fields and CRs.

In this paper, we continue in this direction and address two important questions in particular: (1) what are the physical origins of the diverse Ly$\alpha$ signatures at high $z$: time variation, orientation, or different physical processes? (2) How can we decipher Ly$\alpha$ in terms of the escape of ionizing photons and investigate the role dwarf galaxies played in reionization? For this purpose, we make use of the \textit{PANDORA} simulation suite of dwarf galaxies \citep{Martin-Alvarez23}, which  is one of the first fully coupled radiation-cosmic rays-magneto-hydrodynamic simulation suites of a dwarf galaxy combining a wide range of physical processes. \textit{PANDORA} integrates and encapsulates a complete set of state-of-the-art galaxy formation models and consists of multiple simulation runs with different combinations of relevant physical processes. This allows us to study the dependence of Ly$\alpha$ on different feedback mechanisms. 

Due to the completeness of the simulated physics, this work complements previous Ly$\alpha$ studies by shedding light on how their results might vary with the inclusion of physical processes previously unaccounted for. In this paper, we aim to investigate how Ly$\alpha$ and LyC observables are affected by different physical processes and LOS statistics. This will give us further insights into how Ly$\alpha$ traces the life cycle of dwarf galaxies. We will also compare the results at high  and low redshift, to see whether low-redshift dwarf galaxies are truly analogues of their high-redshift counterparts.

The remainder of the paper is organized as follows. In \autoref{sec:methods} we describe the numerical simulations and post-processing tools we use. In \autoref{sec:case}, we first only focus on exploring the results for the most complete simulation and explore the qualitative properties of the Ly$\alpha$ and LyC emission. We next resort to different simulations in \autoref{sec:diff_phy}, investigating how different physical processes regulate Ly$\alpha$ and LyC emission. In \autoref{sec:discussion} we show how our results relate to the Ly$\alpha$ duty cycle and discuss several caveats associated with our results. We give our conclusions in \autoref{sec:conclusions}.

\section{Methods}
\label{sec:methods}

We here describe the numerical simulations and post-processing techniques we use in this paper. We first give an overview of the \textit{PANDORA} suite of simulations and the implemented physical processes in \autoref{sec:pandora}. In \autoref{sec:lya_rt} we describe how we post-process the simulations with the radiative transfer code \textsc{RASCAS} to generate mock observations. For the rest of the paper, all results are calculated in the galaxy rest frame.

\subsection{Cosmological, radiation-hydrodynamic, zoom-in simulations}
\label{sec:pandora}

We make use of the \textit{PANDORA} cosmological zoom-in simulation suite of a dwarf galaxy \citep{Martin-Alvarez23} and briefly outline the main elements here. The \textit{PANDORA} simulation suite is run with the hydrodynamical code \textsc{RAMSES-RT} \citep{Rosdahl13, Rosdahl&Teyssier15} based on the code \textsc{RAMSES} \citep{Teyssier02}. \textit{PANDORA} resimulates one of the dwarf galaxies described in \citet{Smith19} from $z=127$ to $z=3.5$ and for some models to lower redshift. This dwarf galaxy forms in a halo of virial mass of $\sim 5\times 10^{9} M_\odot$ at $z=3.5$ at the centre of the simulation box of $\sim 14.73$~cMpc on a side. The simulations zoom-in to a central convex hull region about 2.5~cMpc across to track the evolution of the dwarf galaxy and evolve with an advecting passive refinement scalar. The dark matter and stellar particle mass resolution in this region are  $m_{\rm DM}=1.5\times 10^3\,{\rm M}_\odot$ and $m_*=400\, {\rm M}_\odot$. Adaptive Mesh Refinement (AMR) is allowed to reach a maximum resolution of $\Delta x \sim 7$ physical pc. A gas cell is refined when its enclosed mass is larger than 8 $m_{\rm DM}$ or when its size is larger than 4 times the local Jeans length. A metallicity floor of $10^{-4}Z_\odot$ is imposed to allow fragmentation to form Pop II stellar clusters. 

\subsubsection{Gas thermo-chemistry, star formation, supernovae feedback and radiative transfer}

The implementation of thermo-chemistry, star formation, SN feedback and radiative transfer are similar to those of the SPHINX simulation \citep{Rosdahl18} (also have a look at SMUGGLE \citep{Marinacci19, Kannan20} and FIRE \citep{Hopkins14, Hopkins18b, Hopkins23}) and we briefly summarise them here.
The simulations include primordial and metal cooling of hot gas using tables calculated with CLOUDY (above $10^4$ K, \citealt{Ferland98}). For metal cooling of cold gas (below $10^4$K, \citealt{Rosen&Bregman95} ), a piecewise power-law fit to the cooling function is adopted as in \citet{Dalgarno&McCray72}, which take accounts of electronic levels and fine structure levels of neutral and ionized constituents for 7 heavy elements (O, C, N, Si, Fe, Ne, S). 
The star formation prescription adopts a Schmidt law \citep{Schmidt59} $\dot{\rho}_{\rm star} = \epsilon_{\rm ff} \rho_{\rm g}/ t_{\rm ff}$, where $\epsilon_{\rm ff}$ is the star formation efficiency and $t_{\rm ff}$ is the free fall time. Only cells at the highest level of refinement are allowed to form stars \citep{Rasera&Teyssier06}.The cells additionally need to satisfy the condition that gravity outweighs the joint support from turbulent, thermal and magnetic pressure. To model this, the magneto-thermo-turbulent (MTT) star formation model described in \citet{Kimm17, Martin-Alvarez20} is adopted, which inherits the MFFPN model of \citet{Federrath12}. The MTT model calculates star formation efficiency as a function of local gas properties $\epsilon_{\rm ff} (\mathcal{M}, \alpha_{\rm vir} )$, where $\mathcal{M}$ is the Mach number and $\alpha_{\rm vir}$ is the virial parameter. 

The mechanical SN feedback prescription described in \citet{Kimm&Cen14, Kimm15} is employed with a Kroupa initial mass function \citep[IMF;][]{Kroupa01}. This prescription attempts to recover the correct terminal momentum of the snowplough phase of a supernovae remnant regardless of whether the Sedov-Taylor (adiabatic) phase is resolved or not. Each SN injects energy with a specific energy of $\epsilon_{\rm SN}= \frac{10^{51} {\rm ergs} }{10 M_\odot}$ and returns a fraction $\eta_{\rm SN}=0.213$ of exploding mass to the ISM gas, within which a fraction $\eta_{\rm metals}=0.075$ is assumed to be metals. Each star particle triggers its SNe events stochastically sampled from the IMF, with a realistic delay time distribution for SN feedback. A second SN implementation is included, where the feedback strength is boosted by a factor of 4 (Boost; similar to SPHINX implementation)

The radiative transfer is modelled with the implementation in \textsc{RAMSES-RT} \citep{Rosdahl13, Rosdahl&Teyssier15} and in a similar configuration as that in the \textsc{SPHINX} simulation \citep{Rosdahl18}.\footnote{We note that we do not include a homogeneous UVB in these RT simulations.} The coupled radiation hydrodynamic equations are solved with M1 closure to account for photon injection, and radiation-matter interaction via photo-ionization, heating and radiation pressure. The radiation is separated in 3 frequency bins: H~\textsc{i} (13.6 eV $\leq \epsilon_{\rm rad} <$ 24.59 eV), He~\textsc{i} (24.59 eV $\leq \epsilon_{\rm rad} <$ 54.42 eV), and He~\textsc{ii}, (54.42 eV $\leq \epsilon_{\rm rad}$). The explicit treatment of advection usually requires an extremely small time step subject to the Courant condition. To reduce the unnecessary computational cost, RT is subcycled with a maximum of 500 steps over each hydro step and a reduced speed of light of $0.01 c$ is adopted. Stellar particles inject energy into their host cells with a spectral energy distribution (SED) predicted by the BPASSV2.0 model \citep{Eldridge08, Stanway16}, according to their mass, metallicity and age. Regarding the accuracy of the thermal and ionization state of the gas in our simulation. We comment as follows: \citet{Kimm&Cen14} found that a converged escape fraction of ionizing photon requires a resolution of $\sim$ 5 pc (see their Figure 18). Our simulation is therefore marginally convergent since the resolution is $\sim$ 7 pc. The ionization state of the gas is indeed not fully accurately captured with on-the-fly M1-based RT method \citep{Aaron_Smith22}. However, for the warm neutral gas and warm ionised gas that we are mostly interested in ($\gtrsim 10^4$ K), this error is small (see Figure B1 of \citealt{Aaron_Smith22}). 

\subsubsection{MHD and cosmic rays}

A MHD solver based on a constrained transport scheme (CT) \citep{Fromang06, Teyssier06} is used. This algorithm attempts to ensure the solenoidal constraints of the magnetic fields ($\nabla \cdot \vec{B} = 0$) down to numerical precision. The simulations explore two magnetic seeding models: one is a uniform primordial magnetic field \citep{Martin-Alvarez18} and the other is SN-injected magnetic fields \citep{Martin-Alvarez21}. For the latter model, SNe injects small-scale circular loops of magnetic field and 1 percent of their energy as magnetic energy of $E_{\rm inj, mag}=0.01 E_{\rm SN}\sim 10^{49}$~erg. The magnetic field in the simulation is evolved according to the induction equation, and in those models that feature it, through magnetised SN feedback. The average relative divergence in the presence of magnetised SN feedback remains very small $(dx_{\rm cell} \nabla \cdot B / B) < 10^{-9}$ (this magnetised feedback is presented in \citealt{Martin-Alvarez21}; see their Fig.2 for a review of magnetic divergence in simulations featuring this model). During star formation events, the local magnetic field is not modified, and consequently the divergence of the magnetic field remains unchanged.

The CR implementation adopted in \textsc{RAMSES} \citep{Dubois16, Dubois19} accounts for both CR diffusion and streaming. The diffusion coefficient used is $\kappa_\parallel = 3\cdot 10^{28} {\rm cm}^2 {\rm s}^{-1}$, consistent with $\gamma$ ray observations \citep{Ackermann12} and simulations \citep{Salem16}. Beyond diffusion and streaming, hadronic and Coulomb cosmic ray cooling \citep{Guo&Oh08} are also included. The energy source for CR in the simulation is SNe explosion and each SN event injects a CR energy of $E_{\rm CR} = f_{\rm CR}\cdot E_{\rm SN} = 10^{50} {\rm ergs}$. The choice of $f_{\rm CR}=0.1$ is motivated by observations of SN remnants \citep{Morlino&Caprioli12, Helder13}.

\subsubsection{Simulation suite}

The \textit{PANDORA} suite contains a set of simulations with different levels of complexities and physics. Each simulation is named after its included physics and its specific configurations. All simulations contain the default MTT star formation unless indicated otherwise. SN implementation includes both the default and boosted models. The two magnetic field configurations are the astrophysical magnetic field (iMHD) and the strong magnetic field (sMHD). For radiative transfer, we have only one configuration RT. The CR recipes contain full cosmic ray physics (CR) or have CR streaming turned off (nsCR). Here we will focus on the RT, RTiMHD, RTCRiMHD simulations from \textit{PANDORA} in the main text and further discuss the results from RT+Boost, RTsMHD, RTnsCRiMHD simulations in \autoref{sec:other_model} of the appendix (listed in \autoref{tab:fesc_other_model} of this appendix). Interested readers may refer to the original paper for a more detailed description of the physics and a full list of simulations. 

\begin{table}
  \centering
  \begin{tabular}{ccccc}
\hline\hline
Simulation name & RT & SN & Magnetic field & CR \\ \hline
RT & \checkmark & Fiducial & - & - \\
RTiMHD & \checkmark & Fiducial & Astrophysical & - \\
RTCRiMHD & \checkmark & Fiducial & Astrophysical & \checkmark \\
\hline\hline
\end{tabular}

  \caption{Summary of \textit{PANDORA} simulation runs used in this paper. The name convention is the same as \citet{Martin-Alvarez23}.}
  \label{tab:sim_tab}
\end{table}

\subsubsection{Tracking the position of the dwarf galaxy}

In order to obtain accurate Ly$\alpha$ predictions, we require a precise centering for the position of the galaxy. For this, we employ a tracking algorithm for the stellar component. We first characterise the position and properties of the dark matter haloes in the simulations, applying the {\sc HaloMaker} software \citep{Tweed09} to the dark matter component. We identify the system using its precise position at $z = 6$, when it is well-isolated and not undergoing any mergers. The centre of the galaxy is obtained by recursively applying a shrinking spheres algorithm \citep{Power03} solely on its stellar particles, employing the central position of the dark matter halo of interest as an initial guess. To determine the centre of the galaxy for decreasing (increasing) redshift, we re-apply the same algorithm on subsequent (preceding) snapshots. The initial guess employed for each subsequent/preceding snapshot is now the centre of mass of the 500 innermost stellar particles in the previously computed snapshot. Finally, we characterise the galaxy properties (such as the baryonic angular momentum) employing a spherical region with radius $r = 0.2 r_\text{vir}$, where $r_\text{vir}$ is the virial radius of the halo.

\subsection{Synthetic observations of Ly\texorpdfstring{$\alpha$}{alpha} and LyC with \textsc{RASCAS} }
\label{sec:lya_rt}

\subsubsection{Modelling Ly\texorpdfstring{$\alpha$}{alpha} radiative transfer}

Ly$\alpha$ is a resonant line, with its photons undergoing resonant scattering as they propagate through H~\textsc{i} gas. The frequencies and directions of the photons change after each scattering event according to the gas temperature and velocity. Note that this ignores possible sub-grid turbulent velocities which, however, are likely to be small compared to the thermal velocities in warm neutral gas at $10^4$ K (see section 4 in \citealt{Katz22c}). Therefore the emergent Ly$\alpha$ spectra and images are affected by the distribution of both Ly$\alpha$ sources and the cold neutral phase of the ISM and CGM. In this section we describe our method to model these radiative transfer effects.

We model the emission and transport of Ly$\alpha$ photons in post-processing using the publicly available, massively parallel code \textsc{RASCAS} \citep{Michel-Dansac20}. \textsc{RASCAS} adopts a two-step methodology. The first step is to extract from the simulations the information about the medium through which light travels and its radiation sources. The second step performs the radiative transfer and outputs observables.

We extract the computational domain for propagating the photons as a sphere with radius $R$ centered on the galactic centre returned by the tracking algorithm. For the domain, we record information of the gas bulk velocities, number densities of all scatterers ( H~\textsc{i} $\lambda$1216 and D~\textsc{i} $\lambda$ 1215), gas thermal velocities, subgrid turbulent velocities (we assume they are negligible in our model), and dust number densities. We model Ly$\alpha$ sources with three different emission mechanisms - recombination and collisional excitation from gas cells, as well as the stellar continuum near the Ly$\alpha$ wavelength,

\begin{equation}
\begin{aligned}
&\dot{N}_{\mathrm{Ly} \alpha, \mathrm{rec}}=n_{e} n_{\mathrm{H~\textsc{ii}}} \epsilon_{\mathrm{Ly} \alpha}^{\mathrm{B}}(T) \alpha_{\mathrm{B}}(T) \mathrm{d} V, \\
&\dot{N}_{\mathrm{Ly} \alpha, \mathrm{col}}=n_{e} n_{\mathrm{H~\textsc{i}}} C_{\text {Ly } \alpha}(T) \mathrm{d} V. 
\end{aligned}
\end{equation}

For recombination, $n_e$ and $n_{\rm H~\textsc{ii}}$ are the non-equilibrium free electron and ionized hydrogen number densities from the simulations, $\alpha_B(T)$ is the case B recombination coefficient from \citet{Hui&Gnedin97}, and $\epsilon^B_{\rm Ly\alpha} (T)$ is the fraction of recombinations producing Ly$\alpha$ photons from \citet{Cantalupo08}. For collisional excitation, $n_{\rm H~\textsc{i}}$ is the neutral hydrogen number density, and $C_{\rm Ly\alpha} (T)$ is the collisional excitation rate from level 1s to 2p, evaluated using the fit from \citet{Goerdt10}. 

\textsc{RASCAS} samples the real photon population with photon packets. We modify the standard photon sampling strategy to model $N_{\rm MC}$ photons from the inner region and $N_{\rm MC}$ additional photons from the outer region to increase the resolution of photons in the outskirts of galaxies. We adopt $N_{\rm MC}=2 \times 10^5$ for recombination and collisional excitation, respectively\footnote{We use $N_{\rm MC}=1 \times 10^5$ for a few snapshots due to computational cost.}. Since the computational domain is a sphere with radius $R$, we select the inner region to be a sphere with radius $R/4$ and the outer region to be a shell from inner radius $R/4$ to outer radius $R$. The gas-frame frequency of Ly$\alpha$ photon packets from each gas cell sample a Gaussian distribution with the width set by the thermal broadening $\Delta v_{\rm D}$. We ignore the subgrid turbulent motions as they are usually sub-dominant compared to the thermal motions. This frequency is then converted to the external frame based on the gas bulk velocity. The initial directions of photon packets are sampled from an isotropic distribution. 

At each scattering event, Ly$\alpha$ photons change their outgoing direction $\hat{k}_{\rm out}$ relative to the incoming direction $\hat{k}_{\rm in}$ through a phase function $P(\mu)$, where $\mu=\hat{k}_{\rm in}\cdot \hat{k}_{\rm out}$. The phase function depends on the photon's frequency in the frame of scattering $\nu_{\rm scat, in} = \nu_{\rm in} (1- \hat{k}_{\rm in} \cdot \vec{v}_{\rm scat}/c )$. For the scattering of Ly$\alpha$ photons around the core of the line, \textsc{RASCAS} adopts the phase function from \citet{Hamilton40, Dijkstra&Loeb08} and for Ly$\alpha$ photons in the line wings \textsc{RASCAS} adopts Rayleigh scattering. In a high H~\textsc{i} opacity environment, Ly$\alpha$ photons will undergo local scattering multiple times until their frequencies are shifted enough to make a large excursion. To reduce this unnecessary computational overhead, \textsc{RASCAS} adopts a core-skipping algorithm \citep{Smith15} to shift the photons to the line wing without local scattering in space.  We further cap the core-skipping threshold $x_{\rm crit}$ with the maximum value of $x_{\rm crit, max} = 1000$. \textsc{RASCAS} also includes the recoil effect and the transition due to deuterium with an abundance of D/H $= 3 \times 10^{-5}$. Ly$\alpha$ absorption by dust is modelled following \citet{Laursen09b}. The dust density is calculated as,

\begin{equation}
n_{\rm dust}=\frac{Z}{Z_0} (n_{\rm H~\textsc{i}} + f_{\rm ion} n_{\rm H~\textsc{ii}} ),
\end{equation}

where $f_{\rm ion}=0.01$ and $Z_0=0.005$. We use the "Small Magellanic Cloud" (SMC) model described in \citet{Laursen09b} as incorporated in \textsc{RASCAS}. We calculate the dust attenuation $A_V$ following the discussion in \citet{Draine11}. We first calculate the  optical depth in the V band $\tau_V = N_{\rm dust} \sigma_V$, where $\sigma_V$ is the dust cross-section in the V band, calculated as in \citet{Gnedin08}. We then calculate the dust extinction as $A_V = 1.086 \tau_V$.

Ly$\alpha$ photons can either be absorbed or scattered by interaction with dust. The probability of scattering is set by the dust albedo $a_{\rm dust} = 0.32$ following \citet{Li&Draine01}. The scattering angle after each scattering is given by the Henyey-Greenstein phase function \citep{Henyey&Greenstein41} and the asymmetry parameter is set to $g=0.73$ again following \citet{Li&Draine01}.    

We generate synthetic position-position-velocity (PPV) datacubes along arbitrary lines-of-sight (LOS) with dimension $N\times N \times N_{\lambda}$, using the `peeling algorithm' \citep{Yusef-Zadeh84, Wood&Reynold99}  as implemented in \textsc{RASCAS}. The `peeling algorithm' treats each scattering event of the photon packet as a point source and assigns its scattered flux in the observed direction to a bin of the data cube, as described by \citet{Costa22} (also refer to \citealt{Laursen09a, Smith15} for more details). We centre our calculations on the position of the galaxy and generate cubes with a spatial scale of side length $L = R$, where $R$ is the radius of computational domain and ranges from 6 to 10~kpc as redshift decreases. We sample the spatial dimensions with $200 \times 200$ pixels. We adopt 150 spectral bins from 1213.3$\angstrom$ to 1218$\angstrom$, giving a spectral resolution of $0.03\angstrom$. We generate these observables along 108 LOSs for selected starburst phase snapshots and 12 LOSs for other snapshots. 

We define the real escape fraction $f_{\rm esc, Ly\alpha}^{\rm real}$ as the fraction of escaped photons from the galaxy. This quantity does not take into account any observational effect such as PSF, finite resolution or sensitivity. In order to reduce the complexity of our analysis and better understand the Ly$\alpha$ properties in our simulation suite, we omit  these observational effects in this paper and use the symbol $\fesclya$ for $f_{\rm esc, Ly\alpha}^{\rm real}$ from now on. We lastly distinguish angle-averaged and individual LOS $\fesclya$ as below,

\begin{equation}
\begin{aligned}
f_{\rm esc, Ly\alpha}^{\rm aavg} &= \frac{ L_{\rm esc} }{ L_{\rm intr} } , \\
f_{\rm esc, Ly\alpha}^{\rm LOS} &= \frac{ dL_{\rm esc}/d\Omega }{ dL_{\rm intr}/d\Omega } ,
\end{aligned}
\end{equation}

where $L_{\rm intr}$ and $L_{\rm esc}$ are luminosity of the intrinsic and escaped photons.

\subsubsection{Modelling the escape of LyC radiation}

Equipped with \textsc{RASCAS}, we can self-consistently calculate the escape fraction along arbitrary directions by dividing the ionizing flux emanating from the galaxies by the total intrinsic ionizing flux. Although computationally more expensive than ray-tracing methods, this allows us to investigate the LOS dependence of $f_{\rm esc, LyC}$. Ionizing radiation is absorbed by neutral hydrogen, singly ionized helium \citep{Osterbrock&Ferland06}, neutral helium \citep{Yan01} and attenuated by dust \citep{Weingartner&Draine01}. It is safe to ignore absorption by molecular hydrogen as its volume-filling fraction and optical depth are significantly smaller \citep{Kimm19}. As Ly$\alpha$ is only sensitive to H~\textsc{i} and dust rather than He~\textsc{i} and He~\textsc{ii}, we only focus on the LyC photons near the ionization potential of H~\textsc{i} ($850 \angstrom < \lambda < 912 \angstrom$), so our $\fesclyc$ is effectively $f_{\rm esc, 912}$. Similar to $\fesclya$, we also distinguish angle-averaged and individual LOS $\fesclyc$.  

\begin{figure*}
    \begin{subfigure}[b]{\textwidth}
     \centering
     \includegraphics[width=\textwidth]{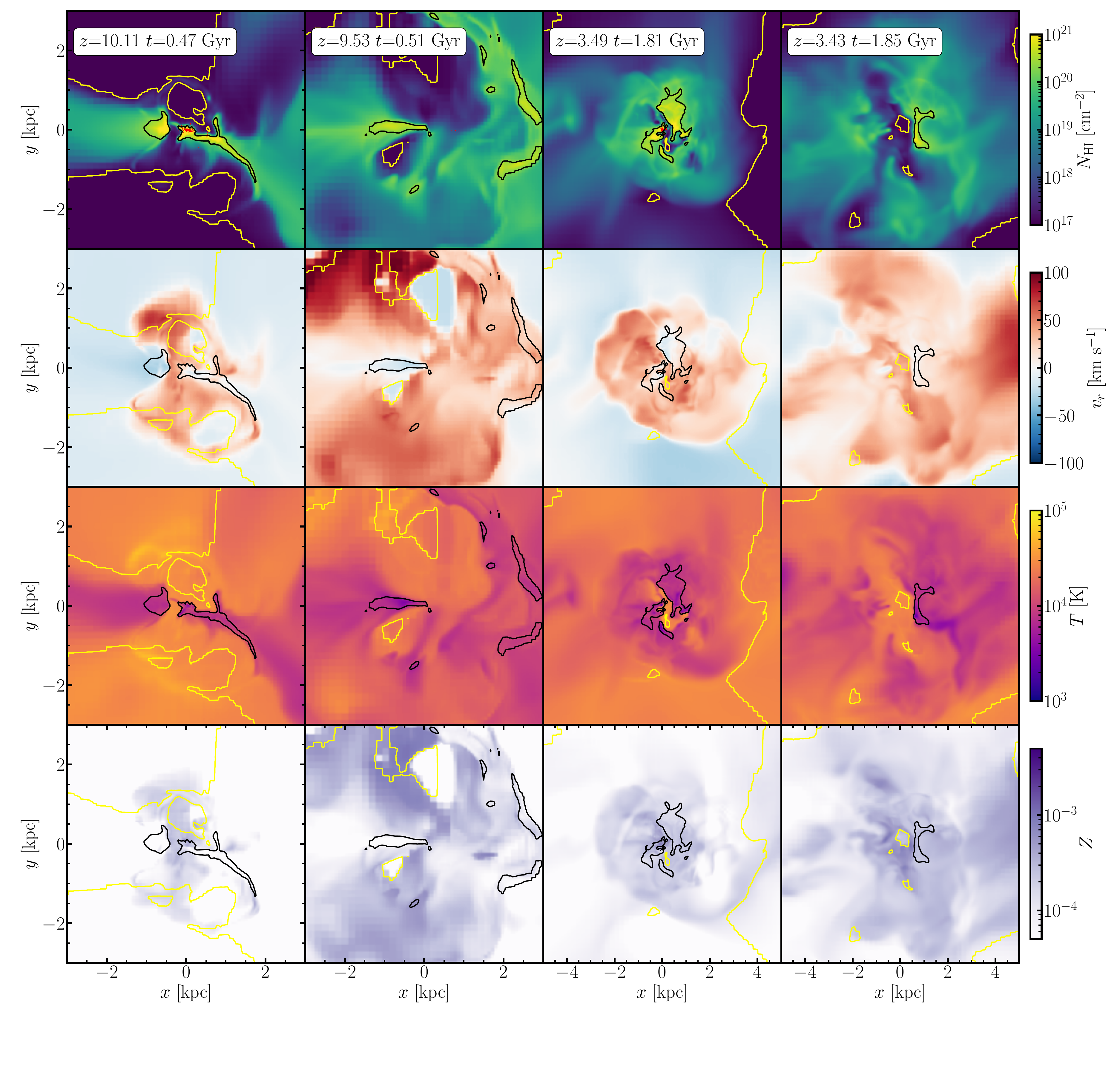}
    \end{subfigure} \\
    \vspace{-1.0cm}
    \begin{subfigure}[b]{\textwidth}
     \centering
     \includegraphics[width=\textwidth]{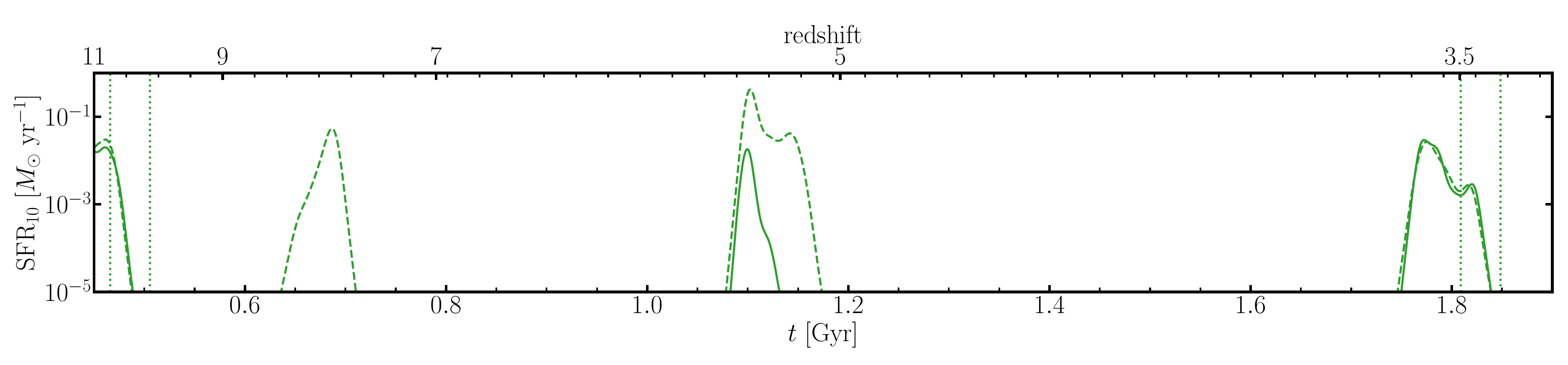}
    \end{subfigure} 
    \caption{
    Top 4$\times$4 panels: Projections of the at-starburst and post-starburst snapshots of the galaxy during its star formation episodes at  `high' (left two columns) and `low' (right two columns) redshifts in the RTCRiMHD simulation. Redshift and time (in Gyr) of each snapshot are indicated in the label in the top panels. From top to bottom the panels show the spatial distribution of H~\textsc{i} column density, and projections of the H~\textsc{i} mass-weighted radial velocity, temperature, and metallicity. The yellow and black contours correspond to $\log N_{\rm H~\textsc{i}} = 17$, and $\log N_{\rm H~\textsc{i}} = 20.3$, respectively. The red dots in the first row represent star particles younger than 30~Myr. The size of the region shown are 6~kpc and 10~kpc at `high' and `low' redshift, respectively. 
    Bottom panel: SFR averaged over 10~Myr as a function of cosmic time, with the times of the snapshots above marked with vertical dotted lines. The solid line shows the in-situ SFR of the galaxy or its main progenitor, while the dashed line shows combined SFR from all the progenitors of the galaxy. For both high and low redshift, the galactic wind is launched at the first of the two selected snapshots, and fully developed in the second. The launching of outflows significantly disperses the surrounding gas and enriches it with metals. }
    \label{fig:general}
\end{figure*}

\begin{figure}
    \centering
    \includegraphics[width=1.\linewidth]{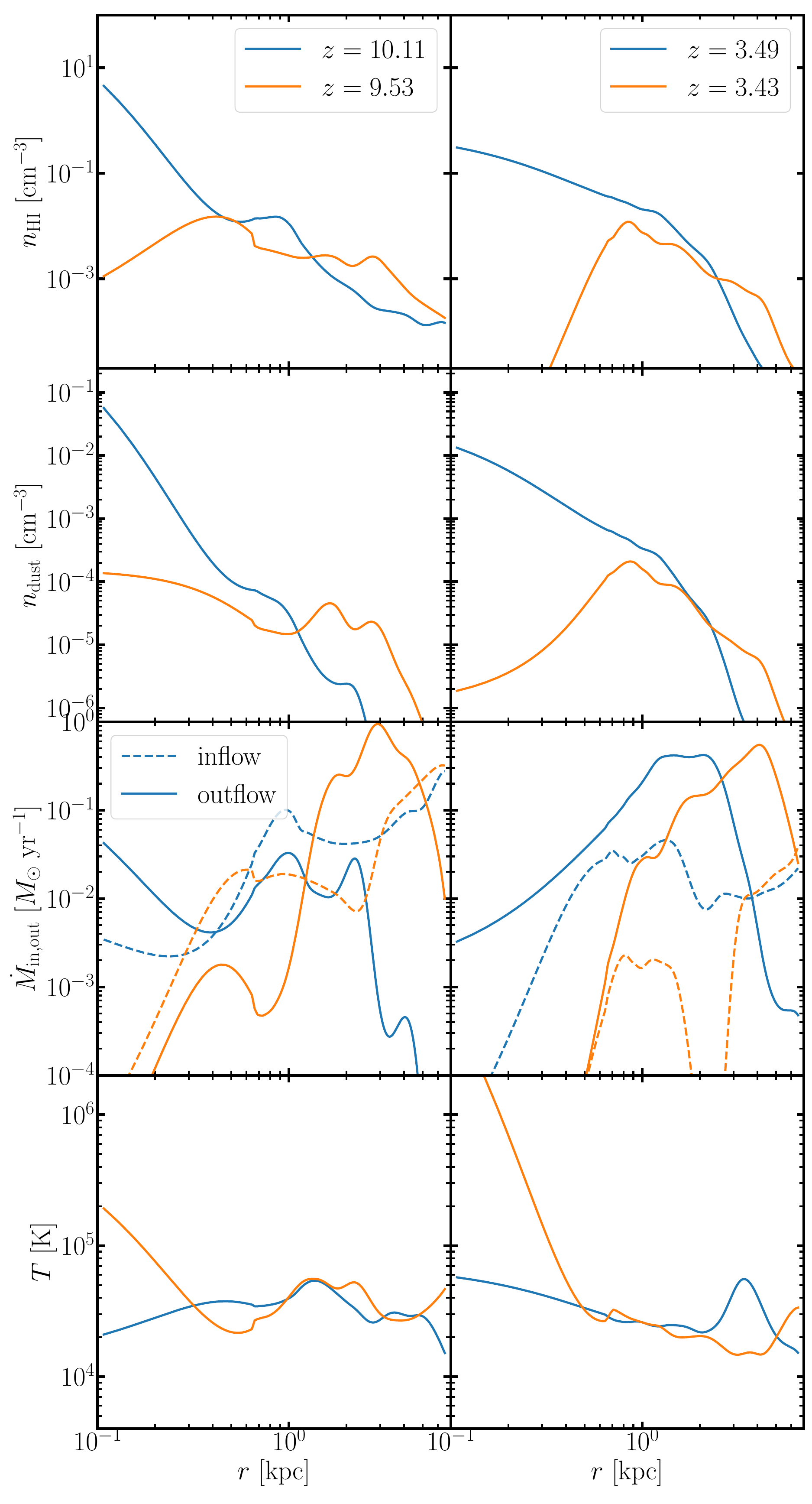}
    \caption{From top to bottom the panels show radial profiles of $n_{\rm H~\textsc{i}}$, $n_{\rm dust}$, H\textsc{i} mass inflow and outflow rate, and mass-weighted temperature, for the at-starburst and post-starburst snapshots at `high' and `low' redshift in the RTCRiMHD simulation. The left column shows the results at `high' redshift while the right column shows those at `low' redshift, as indicated in the label in the top panels. We clearly see the radial expansion of the outflows, the enrichment with dust in the outskirts and the heating of the central cavities during the starburst events. }
    \label{fig:rad_prof_diff_time}
\end{figure}

\subsubsection{Ly\texorpdfstring{$\alpha$}{alpha} observables definitions and measurement}

To perform further quantitative analysis, we define several parameters of Ly$\alpha$ haloes and Ly$\alpha$ spectra in this section.

For Ly$\alpha$ haloes, we define the scale radius to be  
\begin{equation}
    r_{\rm LAH} = \int \mathrm{SB}(r) r dA/\int SB(r) dA,
\end{equation}
where SB is the surface brightness of the Ly$\alpha$ image. Note that here we keep using the galactic centre returned by our tracker algorithm. The Ly$\alpha$ halo broadening factor is defined as the ratio between the scale radius of scattered and intrinsic Ly$\alpha$ halo, $b_{\rm LAH} = r_{\rm LAH}/r_{\rm LAH, intr}$. In real observations, the intrinsic Ly$\alpha$ halo, if dominated by recombination emission, can be measured indirectly through H$\alpha$ and H$\beta$ emission. However, this becomes more complicated if dust scattering and absorption are important.

For scattered Ly$\alpha$, we define the line centre $v_0$ as the first moment of the intrinsic Ly$\alpha$ spectra. We parameterize each peak as an asymmetric Gaussian profile \citep{Shibuya14b},

\begin{equation}
f(\lambda; A, \lambda_0^{\rm asym}, a_{\rm asym}, d)=A \exp \left(\frac{-\left(\lambda-\lambda_0^{\text {asym}}\right)^2}{2 \sigma_{\rm asym}^2}\right),
\end{equation}

where $A$ is the flux intensity and $\lambda_0^{\rm asym}$ is the centre of the peak. $\sigma_{\rm asym} = a_{\rm asym} (\lambda-\lambda_0^{\rm asym})+d$, where $a_{\rm asym}$ is the asymmetric factor and $d$ is the characteristic width of the peak. 

Ly$\alpha$ spectra generally display double-peak features. In some cases, Ly$\alpha$ spectra show a single-peak feature due to a strong inflow or outflow event. These spectra may also show a triple-peak feature due to the presence of deuterium (see \citealt{Dijkstra06}) or more frequently, due to complex gas dynamics.

In order to capture all the cases described, we fit all our Ly$\alpha$ spectra to three models with single-, double-, and triple-asymmetric Gaussian profiles (denoted as s-, d- and t-AG). These are defined as:

\begin{equation}
\begin{aligned}
F_{\rm s-AG} =& f(\lambda; A_1, \lambda_1, a_1, d_1), \\
F_{\rm d-AG} =& f(\lambda; A_1, \lambda_1, a_1, d_1) + f(\lambda; A_2, \lambda_2, a_2, d_2), \\
F_{\rm t-AG} =& f(\lambda; A_1, \lambda_1, a_1, d_1) + f(\lambda; A_2, \lambda_2, a_2, d_2) \\
&+ f(\lambda; A_3, \lambda_3, a_3, d_3).
\end{aligned}
\end{equation}

We compute the likelihood function, $\mathcal{L}$, assuming a SNR of 10 for the three models. We use this likelihood to calculate the Akaike information criteria $\mathrm{AIC} = 2k - 2 \ln \mathcal{L}$, where $k$ is the number of parameters used in each fit (see \citealt{Sharma17}). The model with the minimum value of AIC among all three is selected as the most likely fit, and employed for further analysis. 

We next define line shape parameters. Since a double-peak profile is the most common shape in observations, we define multiple spectral parameters associated with this type of spectrum as follows,

\begin{itemize}
\item velocity separation between blue and red peak, 
\begin{equation}
\qquad \vsep = v(\lambda_2) - v(\lambda_1),
\end{equation}
\item central fraction of spectra within $\Delta v=80$ km s$^{-1}$, 
\begin{equation}
\qquad \fcen = \int_{v_0-\Delta v/2}^{v_0+\Delta v/2} F_{\rm Ly\alpha} dv /\int_{v_{\rm min} }^{v_{\rm max} } F_{\rm Ly\alpha} dv,
\end{equation}
where [$v_{\rm min}$, $v_{\rm max}$] is the Ly$\alpha$ spectral velocity range.
\item velocity width of red peak,
\begin{equation}
\qquad \sigma_{\rm red} = c d_{\rm red} / \lambda_{\rm Ly\alpha},
\end{equation}
\item and red to blue ratio, 
\begin{equation}
\qquad {\rm R2B} = \int_{-\infty}^{v_0} F_{\rm Ly\alpha} dv / \int_{\lambda_0}^{+\infty} F_{\rm Ly\alpha} d\lambda .
\end{equation}

\end{itemize}

The definition of $\vsep$ is modified in the other two cases while R2B and $\fcen$ remain the same. For single-peak profiles we define the velocity separation as $\vsep = 2 v(\lambda_1)$. For triple-peak profiles the situation is more complicated. We find in most cases that there are two peaks on the blue side of the spectrum and a single peak on the red side. We define the wavelength of the blue side $\lambda_{\rm blue}$ to be the weighted average of the wavelengths of the two peaks $(\lambda_1 S_1+\lambda_2 S_2)/(S_1+S_2)$ where $S_i$ is the area of the $i$-th peak. In this scenario, we define the wavelength of the red peak of the spectrum simply as $\lambda_{\rm red}=\lambda_3$. Consequently, we compute the velocity separation $\vsep = v(\lambda_{\rm red}) - v(\lambda_{\rm blue} )$. If there are one peak in the blue part of the spectrum and two peaks on the red side, we define $\vsep$ in an analogous manner. In the extremely rare cases where there are triple peaks on one single side of the spectrum we leave $\vsep$ undefined.

\section{Results: Case study of the RTCRiMHD simulation}
\label{sec:case}

In this section, we explore the results of our Ly$\alpha$ radiative transfer simulations with \textsc{RASCAS}. We first focus on the simulation run with the most complete set of physics, RTCRiMHD. In \autoref{ssec:gas_dist} we investigate the general properties of ISM and CGM that act as a scattering medium for Ly$\alpha$ photons. We then explore the Ly$\alpha$ emission arising in \autoref{ssec:lya_emi} and the variations due to different LOSs in \autoref{ssec:ori_effect}. 

As discussed in \citet{Martin-Alvarez23}, the dwarf galaxy in the RTCRiMHD simulation features the strongest starburst events among the entire suite of \textit{PANDORA} simulations. The dwarf galaxy builds up its stellar mass through the star formation of all its progenitors and the mergers of these progenitors. The main progenitor of the dwarf galaxy undergoes three main star formation episodes triggered by a galaxy merger and gas inflows at $z = 9.5$, 5 and 3.5. The rest of the main progenitor stellar mass build-up is largely gradual. In the analysis below, we focus on the starburst episodes at $z = 9.5$ and $z = 3.5$ which we will denote as `high' and `low' redshift, respectively. 

\subsection{General properties of neutral gas}
\label{ssec:gas_dist}

We show the evolution of the dwarf galaxy during its star formation episodes at both `high' ($z\sim 9.5$) and `low' ($z \sim 3.5$) redshifts in \autoref{fig:general}. The dwarf galaxy undergoes feedback duty cycles at these two redshifts, which impact significantly both the ionization structure of the ISM and the morphology of the galaxy. The panels from top to bottom display the spatial distribution of H~\textsc{i} column density, as well as H~\textsc{i} mass-weighted radial velocity, temperature and metallicity of the dwarf galaxy. The yellow contour highlights column densities $N_{\rm ssh} = 10^{17} {\rm cm}^{-2}$, above which the gas becomes self-shielded and optically thick to the ionizing UVB. LOS with column density above this threshold will be identified as an LLS in QSO absorption spectra. Black contours indicate the minimum column density of DLAs, $N_{\rm DLA} = 10^{20.3} {\rm cm}^{-2}$. Above this value the gas is mainly neutral and creates deep absorption troughs exhibiting damping wings.   

At $z\sim 10$, the continuous inflows trigger the first starburst of the entire simulation run. Young stars form and emit UV photons that ionize the surrounding gas. Subsequent SNe explosions drive a metal enriched hot wind and its ram pressure drives a cold phase outflow. The outflow is not strong enough to extend beyond the virial radius and it interacts with the remaining inflow. After some additional time, at $z=9.53$, enough momentum and energy are deposited into the surrounding ISM for a galactic super-wind to emerge. The super-wind leads to a more clumpy gas distribution, creates multiple low density channels `connecting' the galactic centre to the CGM region, and entrains metals. 

At low $z$, the dwarf galaxy has already built up most of its mass. During its evolution, the dwarf galaxy maintains a large LLS cross section, making it self-shielded from the cosmic UV background. The starburst at low $z$ is merger-triggered rather than inflow-triggered. Before the starburst, two quiescent dwarf galaxies with relatively uniform density structures are approaching each other. They subsequently develop mass transfer between them through a thin dense stream. The two dwarf galaxies start merging at $z\sim 3.8$, and during the merger the galaxy is fed via two external gas streams originating from the CGM. These inflows then trigger star formation activity in the galactic centre at $z=3.5$. Interestingly, we see that the gas morphology after the starburst is somewhat similar to that at high $z$. This is because in the RTCRiMHD simulation although much of the mass of the dwarf galaxy accumulates by $z\sim 3.5$, the star formation activity remains bursty, and when combined with the CR boost to the feedback efficiency, stellar feedback induces a similar impact on the host galaxy compared to high $z$ (see the bottom panel in \autoref{fig:general} ). 

To provide a more quantitative perspective on these results, we show radial profiles of $n_{\rm H~\textsc{i}}$, H\textsc{i} mass inflow and outflow rate, H\textsc{i} gas temperature, $n_{\rm dust}$ in \autoref{fig:rad_prof_diff_time}. When comparing the at-starburst and post-starburst snapshots at both `high' and `low' redshifts, we clearly see the carving out of an ionised core as the galactic wind ejects the neutral gas. This is reflected by a reduction of the inner density and an increase of the density at the outskirts. We see similar trends for the dust density and outflow rates.

\subsection{Ly\texorpdfstring{$\alpha$}{alpha} signature tracing the dwarf galaxy's evolution }
\label{ssec:lya_emi}

\begin{figure*}
    \centering
    \begin{subfigure}[b]{.9\textwidth}
     \centering
     \includegraphics[width=\textwidth]{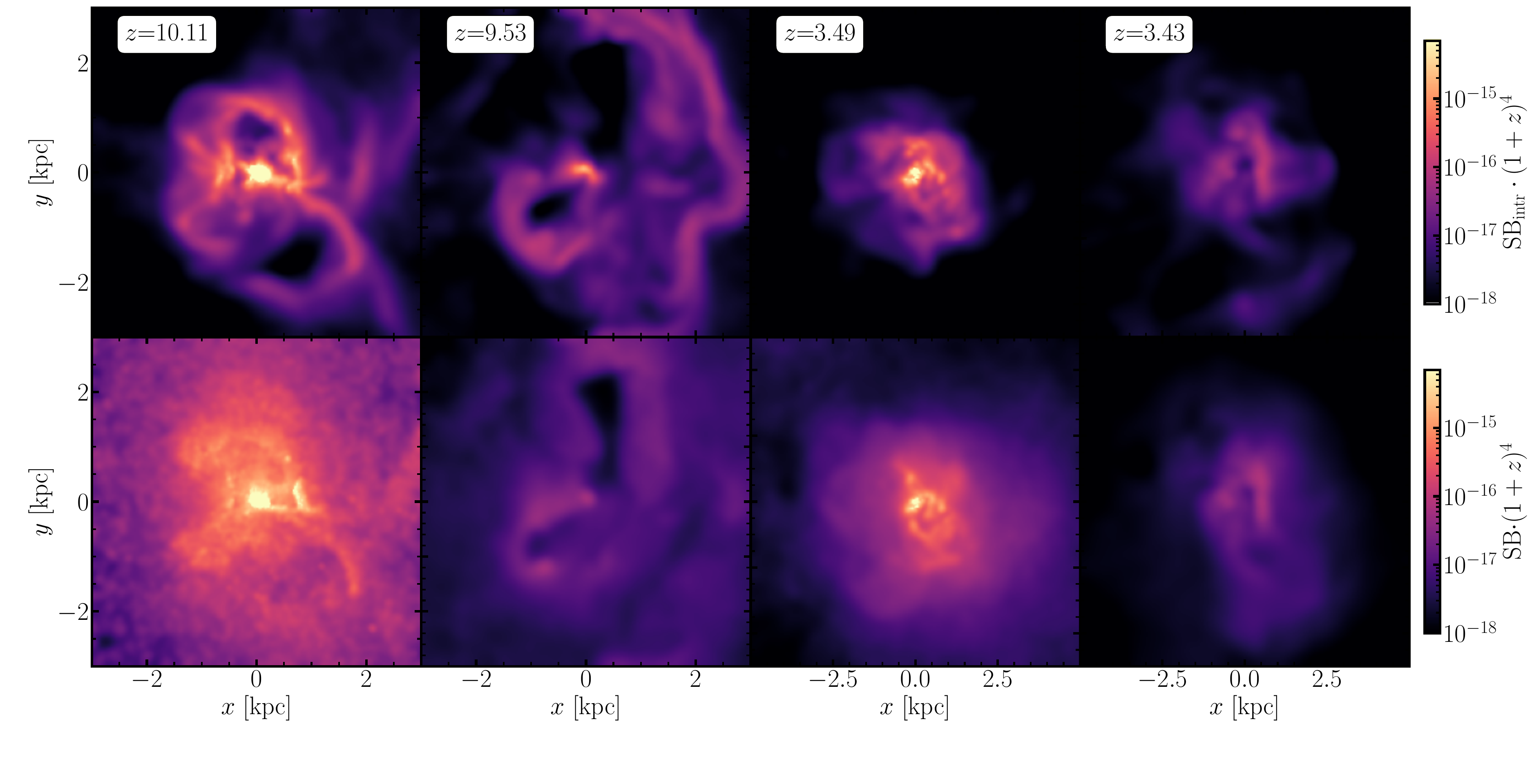}
    \end{subfigure} \\
    \vspace{-0.5cm}
    \begin{subfigure}[b]{.9\textwidth}
     \centering
     \includegraphics[width=\textwidth]{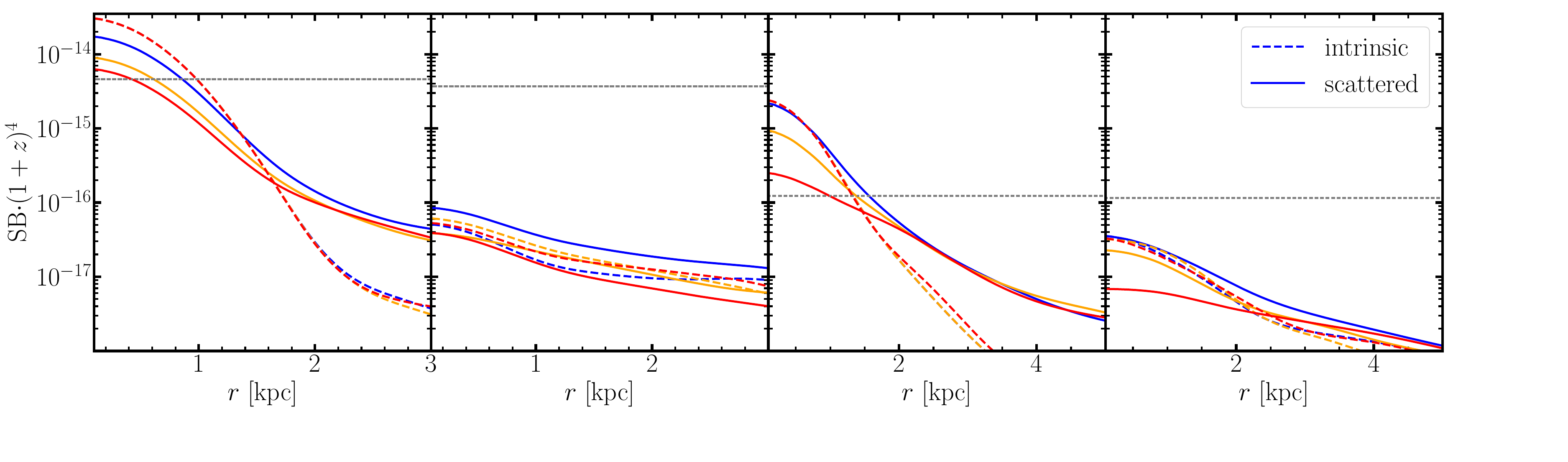}
    \end{subfigure} \\
    \vspace{-0.5cm}
    \begin{subfigure}[b]{.9\textwidth}
     \centering
     \includegraphics[width=\textwidth]{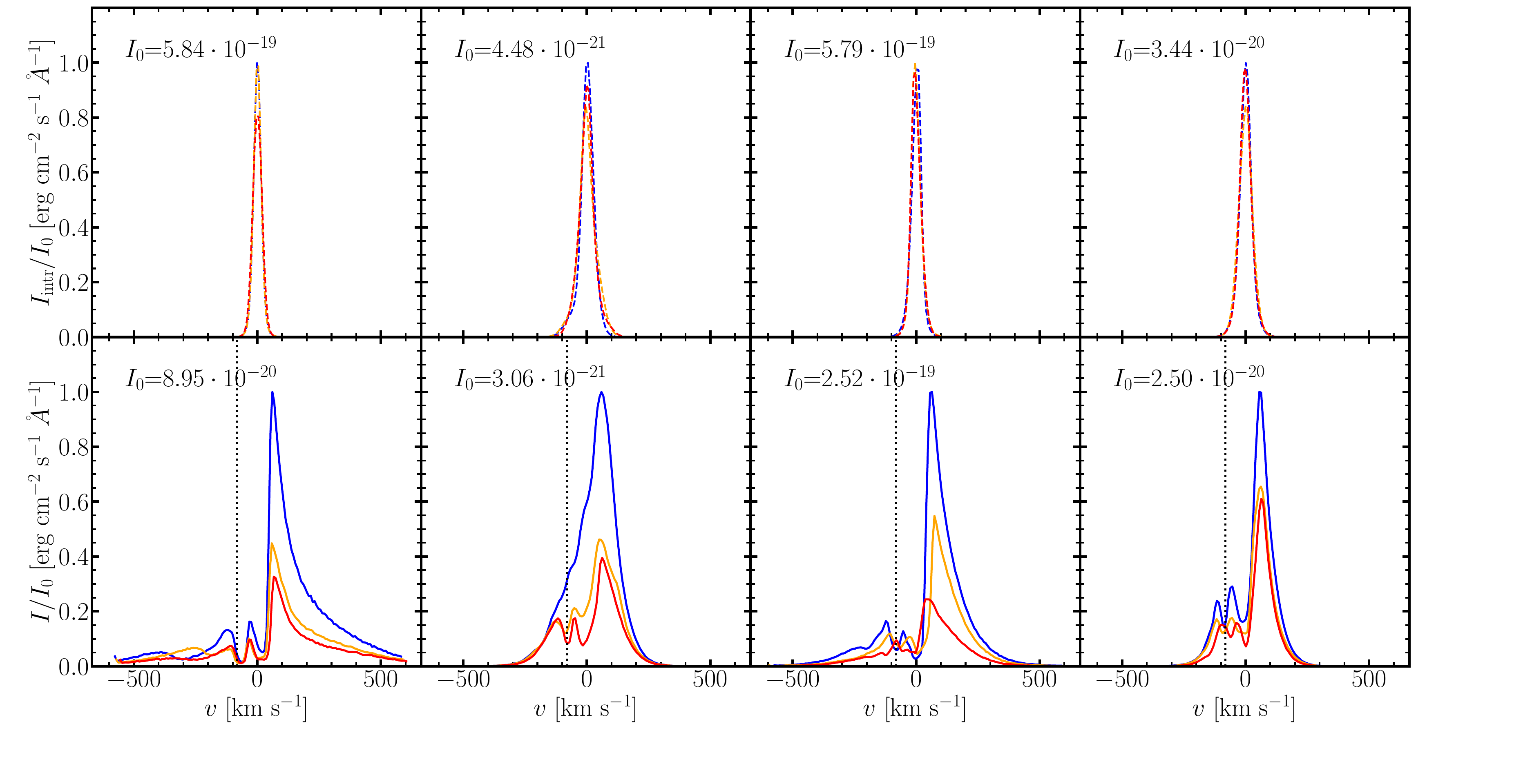}
    \end{subfigure} 
    \caption{Time sequence of Ly$\alpha$ emission for at-starburst and post-starburst snapshots at `high' and `low' redshift in the RTCRiMHD simulation. Each column corresponds to a single snapshot, as indicated in the label in the top panels. The first and second row show the intrinsic and scattered Ly$\alpha$ images. The units for SB$\cdot (1+z)^4$ are erg s$^{-1}$ cm$^{-2}$ arcsec$^{-2}$. These are not included due to lack of space.
    The third row shows the intrinsic (dashed lines) and scattered (solid lines) SB profiles along three different LOS, sampling differences in the scattered Ly$\alpha$ emission. The horizontal dotted line shows the detection threshold $3 \sigma (1+z)^4$ typical of MUSE observations, where $\sigma = 10^{-19}$ erg s$^{-1}$ cm$^{-2}$ arcsec$^{-2}$.
    The fourth and fifth rows show the intrinsic and scattered Ly$\alpha$ spectra for the same three LOS. We have shifted the velocity axis for the Ly$\alpha$ spectra so that the line centre corresponds to zero velocity. The vertical dotted line in the fifth row shows the velocity of the deuterium absorption feature. The deuterium feature makes some of the scattered Ly$\alpha$ spectra show triple-peak profiles.
    Ly$\alpha$ haloes are significantly spatially broadened compared to the corresponding profiles of intrinsic Ly$\alpha$ emission  at the launching of galactic outflows. In contrast, the profiles are only mildly broadened when the outflows are fully developed. During the fully developed outflow stage, Ly$\alpha$ spectra display smaller separations between red and blue peaks, and larger fluxes at zero velocity.}
    \label{fig:lya_diff_time}
\end{figure*}

The line profile of the Ly$\alpha$ emission is very sensitive to the stellar feedback. We now investigate how the outflow properties and evolution are systematically reflected in the Ly$\alpha$ observables of our dwarf galaxy. We show the time evolution of the mock intrinsic and scattered Ly$\alpha$ images, SB profile, and intrinsic and scattered Ly$\alpha$ spectra in \autoref{fig:lya_diff_time}. Comparing the intrinsic and scattered Ly$\alpha$ images, As expected, resonant scattering makes the spatial distribution of Ly$\alpha$ emission significantly more extended than that of the intrinsic Ly$\alpha$ emission. The scattered SB profile is only above the typical MUSE detection threshold, $3 \sigma (1+z)^4$, where $\sigma = 10^{-19}$ erg s$^{-1}$ cm$^{-2}$ arcsec$^{-2}$ (shown by the horizontal dotted lines) within $\sim$ 1-2 kpc at $z=10.11$ and 3.49\footnote{The beam size of MUSE is about 1 arcsec (which corresponds to a size of $\sim$ 7.5 kpc at $z=3.5$). This means  beam smearing effects will smooth out the Ly$\alpha$ image and make our simulated galaxy  too faint to be detected with  current telescopes.}. Intrinsic SB radial profiles show little LOS variations whereas the profiles of scattered Ly$\alpha$ show large LOS variations, especially near the centre. This is because feedback events in the centre create low density channels only along some LOS. Intrinsic spectra show a narrow peak around the line centre with little variation between different LOS. Conversely, most of the scattered spectra show red-peak-dominated double-peaked profiles with significant LOS variations. In general, resonant scattering from dispersed H~\textsc{i} gas is effective at redistributing Ly$\alpha$ photons to larger distances and velocities, and is highly anisotropic consistent with previous findings in \citet{Laursen09a, Barnes10, Barnes11, Smith15}. Some spectra show triple-peak profiles with two peaks on the blue side. The third peak is sometimes due to the extra absorption feature of deuterium (marked as vertical dotted black lines).

We now explore the large variation of the Ly$\alpha$ signatures during the development of the galactic outflow across different redshifts, and defer the investigation of the LOS effect to \autoref{ssec:ori_effect}. At high $z$, the onset of the first starburst ionizes the surrounding gas, generating a large amount of intrinsic Ly$\alpha$ emission. At the time of wind launching, intrinsic Ly$\alpha$ haloes are concentrated, with resonant scattering considerably broadening their light distribution. The spectra show a prominent red peak, as fast outflowing H~\textsc{i} gas is more efficient at scattering photons blueward of Ly$\alpha$. After the galactic wind becomes fully developed and neutral gas has been ejected to large radii, the intrinsic Ly$\alpha$ haloes become more extended while in the centre there is now less recombination emission due to the lower SFR. As a result, resonant scattering results in less broadening than before. This time the spectra show a much clearer blue component, probably because the H~\textsc{i} column density in the wind is larger. The evolution at low $z$ is similar to that at high $z$. 

\begin{figure*}
    \centering
    \includegraphics[width=1.\linewidth]{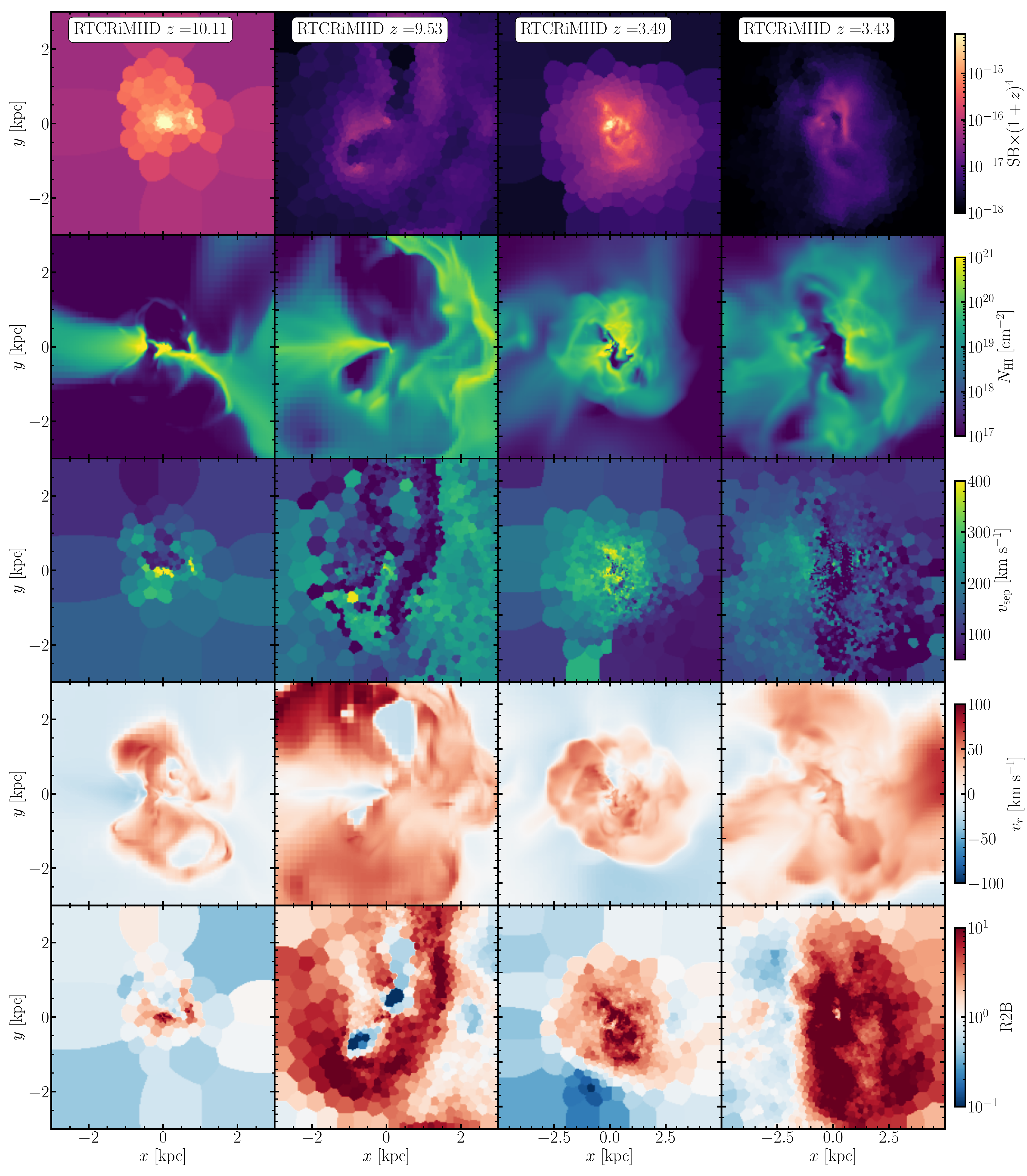}  
    \caption{Time sequence of spatially resolved Ly$\alpha$ spectra, for the at-starburst and post-starburst snapshots at `high' and `low' redshift in the RTCRiMHD simulation. Each column corresponds to a single snapshot, as indicated in the label in the top panels. We show spatially resolved synthetic observation maps of scattered Ly$\alpha$ images, $\vsep$, and R2B (1st, 3rd, 5th rows), and projected maps of H~\textsc{i} column density and H~\textsc{i} mass-weighted $v_r$ (2nd and 4th rows). $\vsep$ maps clearly trace the $N_{\rm HI}$ maps whereas the R2B maps trace the development of outflows, albeit indirectly.}
    \label{fig:res_lya_diff_time}
\end{figure*}

One of the ultimate goals for the study of Ly$\alpha$ is to connect the 3D position-position-velocity Ly$\alpha$ data to the underlying gas kinematics. Although this connection is highly complex, we make a first attempt here by considering spatially resolved Ly$\alpha$ spectral parameter maps. We construct these maps with the following steps: i) we discretise our image into a Voronoi tessellation using the code from \citet{Cappellari&Copin03, Cappellari09} with a SNR of 10; ii) we extract the Ly$\alpha$ spectra for each Voronoi bin from the PPV data; iii), we calculate the line shape parameters (SB, $\vsep$ and R2B) of the spectrum in each bin and form the spatially resolved line parameter map, following a similar pipeline as described in \citet{Erb18, Erb23}. In \autoref{fig:res_lya_diff_time}, we show spatially resolved maps of scattered Ly$\alpha$ images, $\vsep$, and R2B (1st, 3rd, 5th rows), as well as maps of H~\textsc{i} column density and $v_r$ (2nd and 4th rows). 

We first see that our Voronoi tessellation captures the structure of the original scattered Ly$\alpha$ images, with luminous regions sampled with finer bins. These maps provide a sensitive probe of the underlying gas distribution. As expected, the H~\textsc{i} column density correlates with $\vsep$ due to Ly$\alpha$ photons undergoing more local resonant scattering in the largest column densities \citep{Neufeld90}\footnote{ We note that our findings here might mainly apply to the case where the outflow speed is about or below $100$ km s$^{-1}$. This is because for this velocity range the shell model predicts that the $\vsep$ - $\NHI$ relation is relatively unaffected by the outflow speed, while at higher outflow speeds this relation starts to vary (see Figure 2 in \citealt{Verhamme15}).}. $v_{\rm sep}$ decreases with radius, as the optical depths are lower in the outskirts. As the feedback disperses the gas and low-density channels develop (at $z=3.43$), we find lower values for $\vsep$. The red-peak-dominated region expands radially as the outflow develops, reaching larger radii. Considerable scatter exists as spatially resolved Ly$\alpha$ spectra sometimes display complex spectral features which are hard to analyse with an approach that was designed to fit spatially integrated Ly$\alpha$ spectra. 

\subsection{Orientation angle effect of Ly\texorpdfstring{$\alpha$}{alpha} emission and LyC escape}
\label{ssec:ori_effect}

\begin{figure*}
    \centering
    \includegraphics[width=0.82\linewidth]{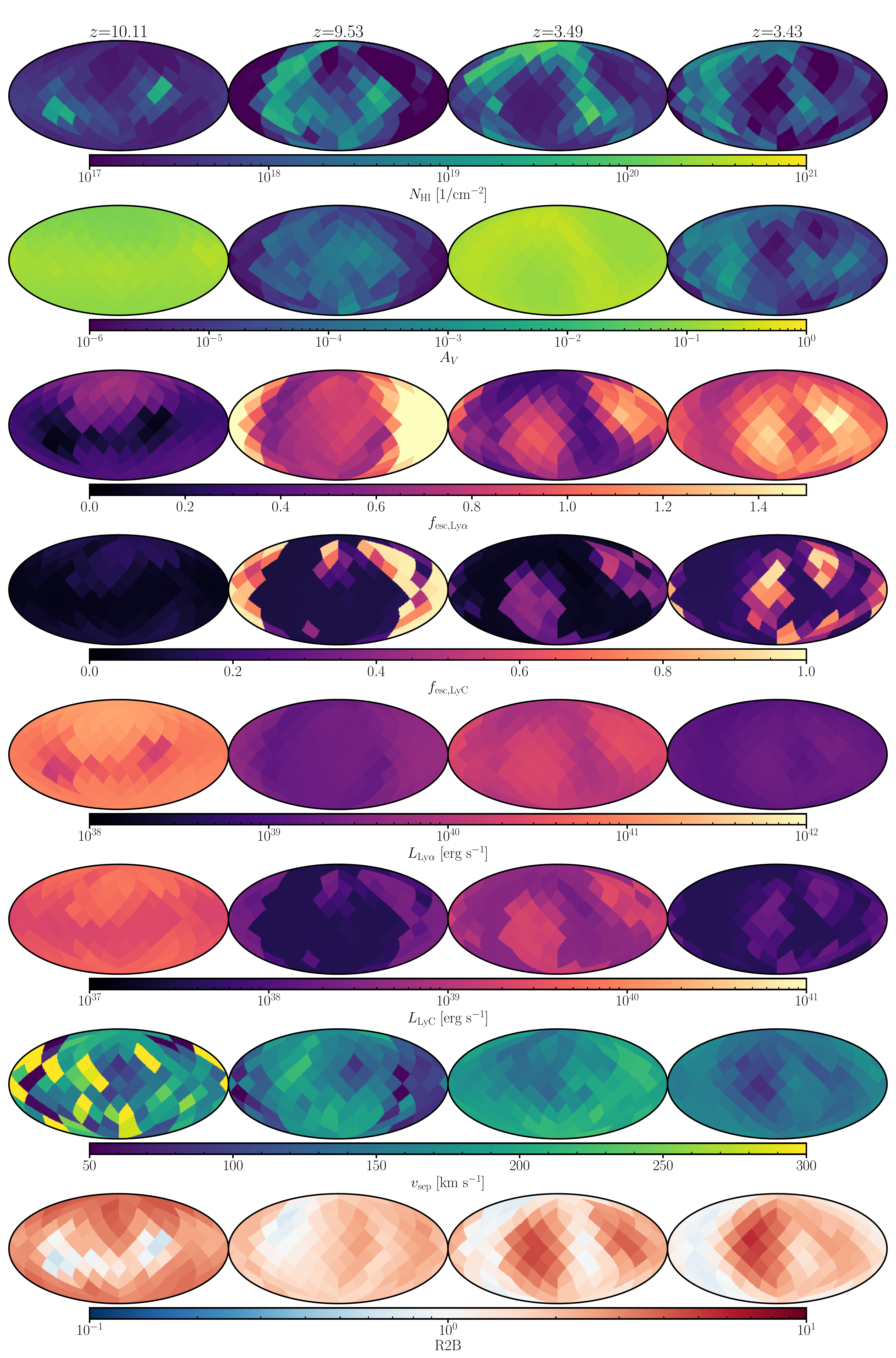} 
    \caption{Time sequence of the angular distribution of various gas properties, Ly$\alpha$ and LyC parameters, for the  at-starburst and post-starburst snapshots at `high' and `low' redshift in the RTCRiMHD simulation. Each column corresponds to a single snapshot, as indicated in the title in the top panels. From top to bottom panels we show full-sky healpix maps centred on the galaxy for $N_{\rm H~\textsc{i}}$, $A_V$, $f_{\rm esc, Ly\alpha}$, $f_{\rm esc, LyC}$, $L_{\rm Ly\alpha}$, $L_{\rm LyC}$, $v_{\rm sep}$, and R2B. The maps show low density channels that have been cleared by feedback. Ly$\alpha$ and LyC have different distributions in and around these channels, indicating different escape mechanisms. Along these channels, the values of and $v_{\rm sep}$ are lower whereas those for R2B are higher.}
    \label{fig:ang_dist}
\end{figure*}

Ly$\alpha$ transport is a highly anisotropic process. As a result, the morphology and spectra of the emerging Ly$\alpha$ haloes depend strongly on the LOS we observe. We explore LOS statistics by plotting the angular distribution of $\NHI$, $A_V$, $\fesclya$\footnote{Note that although the angle-averaged $f_{\rm esc, Ly\alpha}$ is strictly less than 1, individual LOS $f_{\rm esc, Ly\alpha}$ can be larger than 1 due to its highly anisotropic nature.}, $f_{\rm esc, LyC}$, $b_{\rm LAH}$, $\vsep$, and R2B in \autoref{fig:ang_dist}. The full-sky maps are centred on the galaxy and employ a healpix discretization. The $\NHI$ panels show a distribution where feedback processes have cleared out several low density channels (for example, there are two escape channels at $z=3.49$) with high $f_{\rm esc, LyC}$ and $f_{\rm esc, Ly\alpha}$. This indicates a strong correlation between these two quantities, consistent with the results of \citet{Maji22}. However, this correlation is non-trivial, as we see the channel width of $\fesclyc$ is smaller than that of $\fesclya$ and $\fesclyc$ declines more rapidly towards the edges of escape channels than $\fesclya$ (cf. Figure 9 in \citealt{Aaron_Smith22}). These differences are due to different escape mechanisms of Ly$\alpha$ and LyC photons, which will discuss below. The dust attenuation $A_V$ is close to 1 during the starburst and declines rapidly to 10$^{-4}$ after the starburst. This means that the dust has an impact on Ly$\alpha$ and LyC emission only during the starburst. Furthermore, we find lower values for $\vsep$, and higher values for R2B along these channels. The values along these channels reflect the low H~\textsc{i} column density and high outflow speed. Interestingly, the R2B distributions mimic that of $\fesclya$, while the $\vsep$ distributions resemble $\fesclyc$ more closely.

\begin{figure*}
    \centering
    \includegraphics[width=1.\linewidth,height=1.\textheight,keepaspectratio]{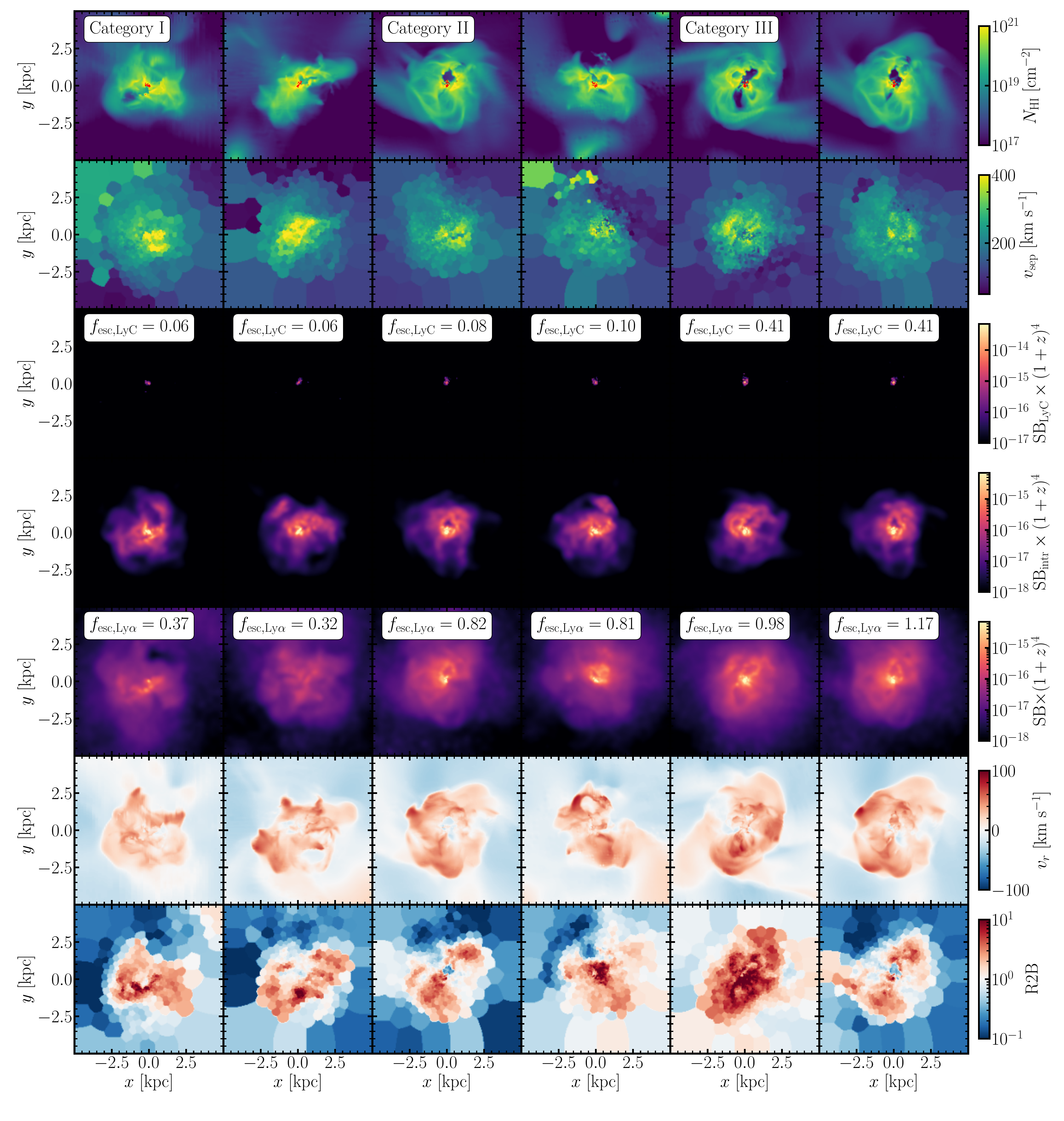}
    \caption{LOS variation of  the dwarf galaxy at $z=3.49$ for the RTCRiMHD simulation. Each column corresponds to a different LOS. The first and second rows show H~\textsc{i} column density map and spatially resolved maps of $v_{\rm sep}$. The red dots in the first row represent star particles younger than 30~Myr. The third row shows LyC images, and the value of $\fesclyc$ in the inset label. The fourth and fifth rows show intrinsic and scattered Ly$\alpha$ images. The labels included inside the panels of the scattered Ly$\alpha$ images indicate the value of $\fesclya$. The sixth and seventh rows show  projections of H~\textsc{i} mass-weighted radial velocity and spatially resolved maps of R2B. 
    We classify these LOS into three categories depending on the values of $\fesclya$ and $\fesclyc$. These are (I; leftmost two columns) low $\fesclyc$ and low $\fesclya$, (II; central two columns) low $\fesclyc$ and high $\fesclya$, (III; rightmost two columns) high $\fesclyc$ and high $\fesclya$. These three categories distinguish whether we can observe young massive stars
    through a low-density channel or not: complete coverage by high column density H~\textsc{i} (I), partial leakage through a low column channel (II) and complete leakage through a low column channel (III). Ly$\alpha$ halo morphologies display considerable variation between each of the different sightline categories.}
    \label{fig:ori_effect}
\end{figure*}

To structure the discussion of  how Ly$\alpha$ and LyC photons travel along different LOSs, we classify all 108 sightlines into three categories (I) low $\fesclya$ (<40\%) and low $\fesclyc$ (<10\%) (II) high $\fesclya$ (>80\%) and low $\fesclyc$ (<10\%) (III) high $\fesclya$ (>80\%) and high $\fesclyc$ (>40\%). We illustrate these categories with 6 sample sightlines (two for each category) in \autoref{fig:ori_effect}. In general, the intrinsic Ly$\alpha$ images all have similar sizes and shapes but the scattered Ly$\alpha$ haloes show different morphologies along different LOSs. This suggests that due to the inhomogeneous distribution of H~\textsc{i} gas, the Ly$\alpha$ photons get scattered anisotropically and give rise to different levels of spatial broadening along different LOS. This spatial redistribution of Ly$\alpha$ photons is fundamentally different than the transport of the photons of the nebular emission lines such as [O \textsc{iii}], giving a natural explanation for the commonly observed spatial offsets between Ly$\alpha$ haloes and nebular line haloes \citep{Williams14, Capak15}. 

We next investigate the difference between the three categories of sightlines. The first set of two columns from the left show category I sightlines. From the extremely low $\fesclyc$, we see young stars residing in the galactic centre are completely covered by high  column densities of  foreground H~\textsc{i}. Ly$\alpha$ photons undergo significant resonant scattering in the centre and escape at the outskirts, producing Ly$\alpha$ haloes with extended morphologies. The second set of two columns shows the category II sightlines, which display more disturbed H~\textsc{i}  morphologies and some poorly resolved low-density channels. Despite these, the stars do not line up with these low density regions and are still covered by dense gas along the LOS. Ly$\alpha$ photons can be scattered to these channels and escape efficiently, showing a more concentrated Ly$\alpha$ halo and high $\fesclya$. The third set of two columns shows the category III sightlines. For this category, the LOS is aligned with the low density channels down to the centre of the galaxy resulting in relatively low H~\textsc{i} column densities, and both LyC and Ly$\alpha$ photons escape efficiently from the centre. 

Comparing these three categories of sightlines, LyC and Ly$\alpha$ present different escape fractions due to their different propagation mechanisms within the galaxy. LyC radiation can only be scattered by dust and otherwise travels in straight trajectories. Because of this, if the stellar distribution is compact, LyC is easily obscured by a small column of foreground gas. Ly$\alpha$ photons, on the other hand, can escape via the so-called ``Neufield mechanism'': they travel in a zig-zag trajectory until they reach low-density channels through which they can escape the ISM. We propose three conditions for the existence of category II of LOSs: (i) compact stellar distribution, (ii) inhomogeneous distribution of gas, (iii) Ly$\alpha$ photon escapes via the Neufield mechanism.

Finally, we compare the gas distribution to the spatially resolved Ly$\alpha$ profile along different directions. We find once again the $\vsep$ maps correlate with the $\NHI$ maps, albeit weakly,  whereas the R2B maps reflect the spatial distribution of outflows.  Interestingly the $\vsep$ maps, given sufficiently high resolution, trace the low density channels near the centre along the category III sightlines. Together with the results discussed above, we conclude that the spatially resolved Ly$\alpha$ profiles reflect the underlying spatial distribution and dynamical state of H~\textsc{i}   along different LOS. 

\section{Results: Variation of Ly\texorpdfstring{$\alpha$}{alpha} signatures with different physical processes}
\label{sec:diff_phy}

We now expand our simulation sample by turning on and off different physical processes with the objective of exploring how they affect the Ly$\alpha$ signatures. We focus on the RT, RTiMHD and RTCRiMHD simulations  in our simulation suite. As we have shown, the emergent Ly$\alpha$ emission is highly affected by the ISM and CGM structure. This means Ly$\alpha$ emission can be a sensitive probe of physics regulating gas distribution and kinematics at ISM scales. In this section, we investigate this systematically by analyzing simulation runs with different complexity and multiplicity of physics ingredients. In particular, we explore the effects of magnetic fields and CRs. We note that getting robust answers to this problem is extremely challenging, as variations due to different physics may be degenerate with the effects of LOS and time variations we studied above. We therefore tackle this problem with two different strategies: (i) carefully select a sample of galaxies in a similar dynamical state (inflow, outflow or `quasi-static') and stellar mass. (ii) perform a large statistical analysis of Ly$\alpha$ on the parameter space spanned by different times, LOSs, and different physics. Our first strategy is designed to investigate the dependence of Ly$\alpha$ observables on different physics, shedding light on how we should design the next-generation galaxy formation models. Our second strategy predicts the variations for observations under a wide range of circumstances, which will assist the interpretation of various current observations and provide valuable insight for future observations. 

\subsection{Time evolution of dwarf galaxies in three simulations}
\label{ssec:time_evo}

\begin{figure*}
    \centering
    \includegraphics[width=1.\linewidth]{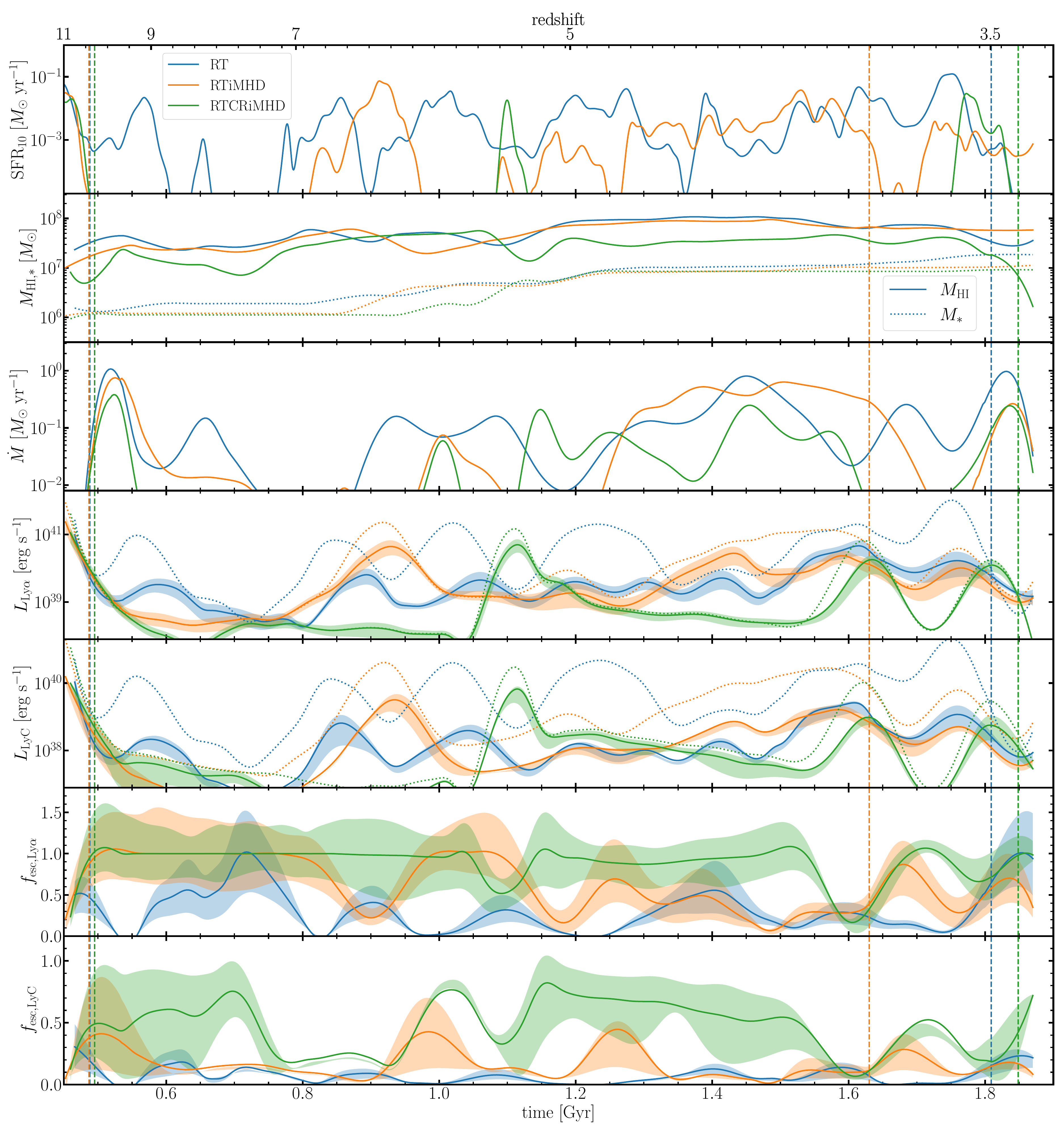}
    \caption{Time evolution of the in-situ SFR averaged over 10~Myr (first row), $M_{\rm HI}$ and $M_*$ (second row), and mass outflow rate (third row). The fourth panel shows the intrinsic Ly$\alpha$ luminosity as dotted lines and angle-averaged values as solid lines. The shaded regions show 16th-84th percentile scattered Ly$\alpha$ luminosities along different LOS. The fifth row is the same as the fourth row, but shows the LyC luminosity. The sixth row shows the angle-averaged values of $\fesclya$ as solid lines and 16-84th percentile values as shaded regions. The seventh row is the same as the sixth row, but shows the results for $\fesclyc$. The set of vertical lines marks the selected snapshots of starburst episodes at both `high' and `low' redshifts of the three simulations with different physics included. These selected snapshots are used for the sample study in \autoref{ssec:sample}. We see that RT has a cyclic SFH, while RTCRiMHD has three distinct starbursts across its evolution. RTiMHD sits somewhere in between. The differences in SFH drive the different evolution of galaxy mass, outflow, as well as the observable properties of Ly$\alpha$ and LyC.}
    \label{fig:time_evo}
\end{figure*}

We show the time evolution of the galaxy's in-situ SFR, H~\textsc{i} mass ($M_*$ mass), H~\textsc{i} mass outflow rate, intrinsic Ly$\alpha$ luminosity, intrinsic ionizing luminosity, and $\fesclya$ and $\fesclyc$ for the three simulation runs in \autoref{fig:time_evo}. The SFH and feedback duty cycle of these three simulations are significantly different. In RT, the dwarf galaxy only undergoes an initial starburst and then settles into a self-regulated SFH at $z \sim 7$ while in RTiMHD the dwarf galaxy undergoes two starbursts and enters into self-regulation at a lower redshift, $z\sim 5$. In the RTCRiMHD simulation, the SFH history of the dwarf galaxy has three distinct and intense starbursting events. Each intense starburst ejects most of the gas and the dwarf galaxy later on slowly accumulates gas with negligible star formation. Note that the RTCRiMHD simulation does not become self-regulated. There are further properties to note when we compare these galaxies at the same redshift. At high $z$, in RT the dwarf galaxy shows extended star formation while in RTiMHD and RTCRiMHD (also in RTnsCRiMHD, which is not included in this plot) the dwarf galaxy undergoes very similar intense starbursts. This suggests that the inclusion of magnetic fields is responsible for the different star formation history  in the earliest phases of this galaxy. At low $z$, both RT and RTiMHD have similar cyclic star formation while RTCRiMHD remains bursty, showing that CR feedback is more effective at lower redshift in our simulations\footnote{Note that although the final stellar masses of the three simulations are similar, the integrals of the SFH are not, especially for RTCRiMHD. We note that this is not contradictory as in RTCRiMHD the stellar mass grows mainly through mergers with other galaxies with already formed stars (not shown in the SFH), rather than in-situ star formation.}.

The pattern of the SFH is the key driver of the Ly$\alpha$ and LyC evolution of galaxies. We first see that both intrinsic Ly$\alpha$ and LyC luminosities scale with SFR and are almost synchronous with the SFH\footnote{The RTCRiMHD simulation still shows a small amount of $L_{\rm Ly\alpha}$ when there is no ongoing star formation. This is because stars in neighbouring galaxies photo-ionise the gas in our galaxy. This gas subsequently produces Ly$\alpha$ photons.}. The evolution of escaped Ly$\alpha$ and LyC luminosities still roughly follow this trend, but are more complicated as they have extra dependencies on their escape fraction.  Since most of the observed LAE has a Ly$\alpha$ luminosity of about $10^{41.5}$~erg s$^{-1}$, our dwarf galaxy is below the detection threshold most of the time.  The evolution of $\fesclyc$ is more complicated due to the feedback requiring some time to create low-density channels and foster the escape of Ly$\alpha$ and LyC photons\footnote{Note that at some snapshots angle-averaged $\fesclyc$ values may be larger than the 84th percentile value of $\fesclyc$ along different LOS. This is because of two reasons: (1) Those snapshots only have 12 LOS so they may not be representative enough. (2) Due to the large LOS anisotropy of $\fesclyc$, its angle-averaged values can be larger than its 84th percentile value.}. Consequently, there is always a delay between the peaks of SFR and $\fesclyc$ \citep{Trebitsch17, Kimm17, Kimm19, Kimm22}. 

Interestingly, the simulations with different physics included present significantly different LyC escape histories. RT always has a relatively low $\fesclyc$ while RTiMHD has several short bursts of LyC leaking phases with $\fesclyc \sim 0.4$. RTCRiMHD features several considerably extended periods of LyC leaking phase with $\fesclyc \sim 0.7$ as each feedback event disperses the gas within the dwarf galaxy almost entirely. The evolution of $\fesclya$ is not only affected by the H~\textsc{i} distribution but also strongly regulated by the dust content.\footnote{The way these two factors together affect $\fesclya$ is non-trivial. Ly$\alpha$ photons can only be absorbed by dust, but the resonant scattering with H~\textsc{i} increases the photon travel length and therefore increases the probability of Ly$\alpha$ photons encountering dust. However, if there is little dust around the galaxy then $\fesclya$ will be close to 1. } Consequently, its evolution is somewhat less dominated by the factors described earlier. This is especially the case for the RTCRiMHD simulation, which features $\fesclya \sim 1$ most of the time due to its very low dust content (with $A_V \lesssim 1$, see second row in \autoref{fig:ang_dist}). We emphasise again that $\fesclya$ here does not take any observational effects into account. Better predicting this quantity requires a better dust model and detailed and careful treatment of observational effects. We will leave this for future work.

We next select snapshots in simulations for our sample studies in \autoref{ssec:sample} below. The objective of the snapshot selection is to study the dwarf galaxy during periods of similar dynamical state and to have comparable H~\textsc{i} and stellar masses. Such a selection will allow us to disentangle time variation and dynamical effects from those due to accounting for different physical processes. For our sample studies we selected snapshots in the fully developed outflow phase at both `high' and `low' redshifts, respectively. The motivation for specifically selecting such snapshots is that they have a higher Ly$\alpha$/LyC escape fraction, luminosity, and hence observability. The RTCRiMHD only has one distinct starburst event each at `high' and `low' redshift while RT and RTiMHD in general have a smoother SFH. For this reason, we first select the outflow snapshots in RTCRiMHD at `high' and `low' redshift, and then in RT and RTiMHD we choose the outflow snapshots that happened at similar times. We select these snapshots with a time lag after the time when the peak of the star formation happens. This time lag allows the dwarf galaxies to fully develop the low-density channels and the extended $\sim$ kpc-scale outflow structures. The exact time length depends on different simulated physics and evolutionary epochs of the galaxies. In general, this time lag is about 20~Myr at `high' redshift and 80~Myr at `low' redshift. We mark the snapshots we select as vertical dashed lines in \autoref{fig:time_evo}. We lastly note that choosing the snapshots at exactly the same dynamical state among different simulation runs is virtually impossible, especially when considering the discrete time sampling of available snapshots that does not fully resolve the feedback duty cycle. 

\subsection{Sample studies for three simulation runs}
\label{ssec:sample}

\begin{figure*}
    \centering
    \includegraphics[width=1.\linewidth]{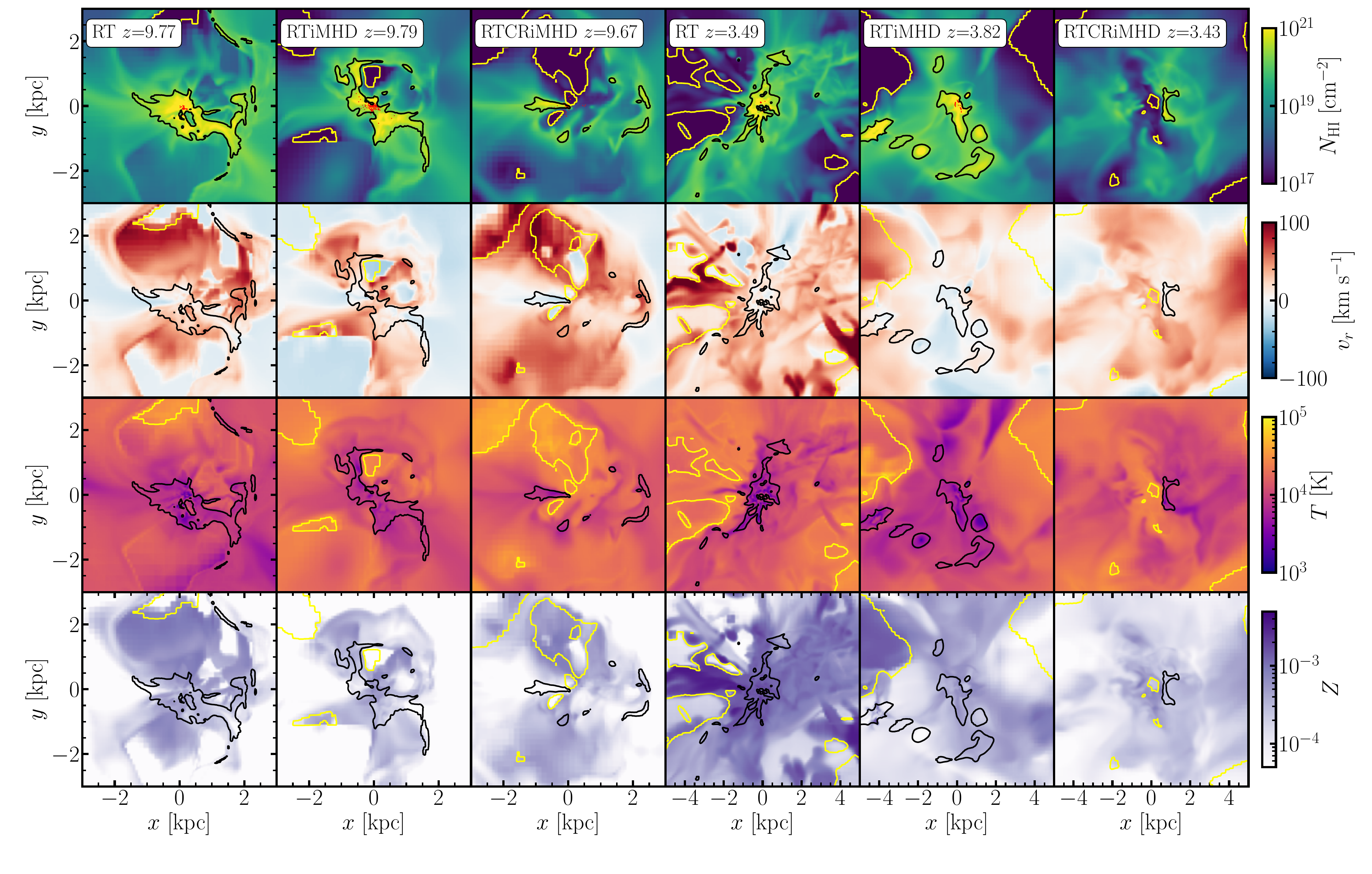}
    \caption{Variations of the properties of the dwarf galaxy due to different simulated physics.
    From top to bottom the panels show the spatial distribution of H~\textsc{i} column density, and projections of the H~\textsc{i} mass-weighted radial velocity, temperature, and metallicity. 
    Different columns correspond to different snapshots and simulations, as indicated in the labels in the top panels.
    The left three columns show the results at `high' $z$ and the right three columns show the results at `low' $z$. In each redshift bin, the three columns show the results of post-starburst snapshots for the RT, RTiMHD, and RTCRiMHD simulations, when the outflows are fully developed.}
    \label{fig:diff_phys}
\end{figure*}

We explore our first strategy in this section. We investigate how gas and Ly$\alpha$ properties vary with different physics and whether our conclusions in \autoref{sec:case} are robust when studying different physical models. 

In \autoref{fig:diff_phys} we show the H~\textsc{i} column density, radial velocity, temperature and metallicity maps of selected snapshots in the RT, RTiMHD, and RTCRiMHD simulations, at both `high' and `low' redshift. We first focus on high $z$. In the RT simulation, the pre-SN feedback,  mainly photoionization-heating, reduces the strength of SNe feedback by reducing the spatial clustering of the SNe \citep{Smith21b}.\footnote{A more detailed analysis of this is left for Martin-Alvarez (in prep).} In RTiMHD, chunks of cold-phase gas are entrained in the CGM probably due to the protection by magnetic fields, especially at low $z$. In particular, we do not see any low-density channel created. This means RTiMHD features the weakest feedback strength among all three runs. In RTCRiMHD with streaming CR, we see the strongest feedback across our entire simulation suite \citep{Martin-Alvarez23}. This results in a much smaller DLA cross section and several low density channels are created. The trends at low $z$ are similar, but RT has a stronger outflow event than at high $z$. This is because RT has the highest SFR at this epoch compared to the other two runs. 

\begin{figure*}
    \centering
    \begin{subfigure}[b]{\textwidth}
     \centering
     \includegraphics[width=\textwidth]{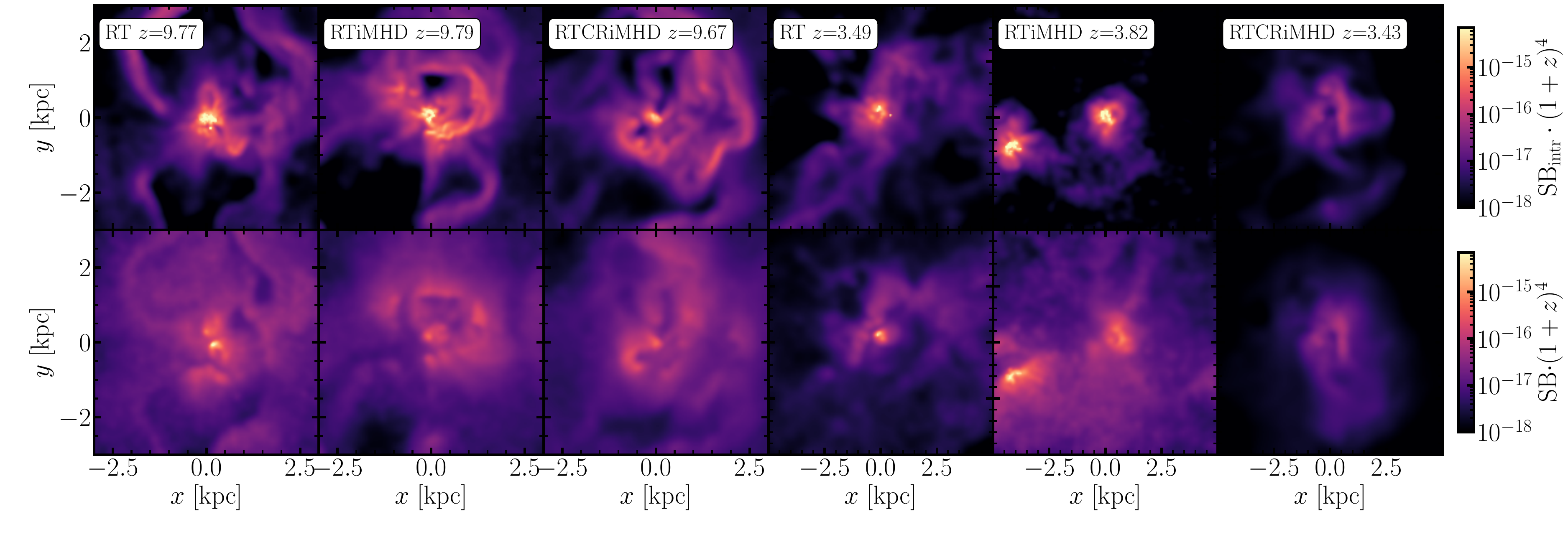}
    \end{subfigure} \\
    \begin{subfigure}[b]{\textwidth}
     \centering
     \includegraphics[width=\textwidth]{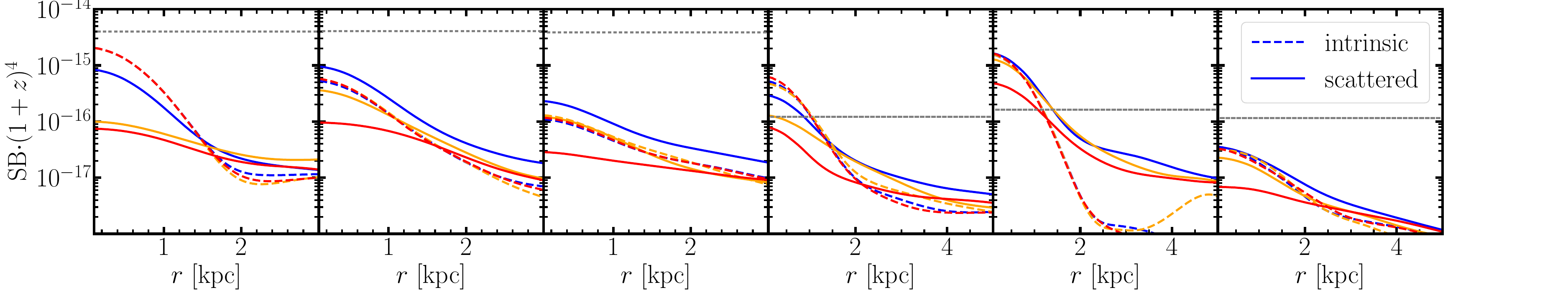}
    \end{subfigure} \\
    \begin{subfigure}[b]{\textwidth}
     \centering
     \includegraphics[width=\textwidth]{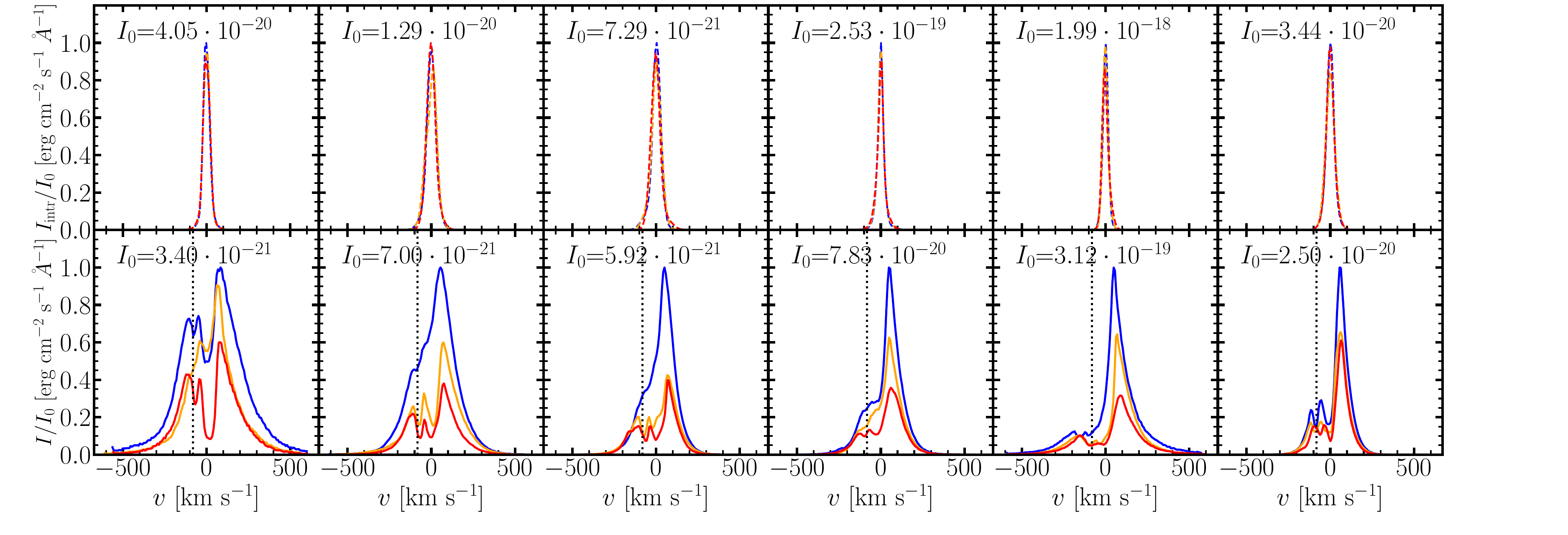}
    \end{subfigure} 
    \caption{Variations of the Ly$\alpha$ observations due to different simulated physics. The first and second row show the intrinsic and scattered Ly$\alpha$ images. The third row shows the intrinsic (dashed lines) and scattered (solid lines) SB profiles along three different LOS sampling differences in the scattered Ly$\alpha$ emission. The fourth and fifth rows show the intrinsic and scattered Ly$\alpha$ spectra for the same three LOSs. 
    Different columns correspond to different snapshots and simulations, as indicated in the labels in the top panels.
    The left three columns show the results at high $z$ and the right three columns show the results at low $z$. In each redshift bin, the three columns show the results of post-starburst snapshots for the RT, RTiMHD, and RTCRiMHD simulations.}
    \label{fig:lya_diff_phys}
\end{figure*}

\autoref{fig:lya_diff_phys} shows intrinsic and scattered Ly$\alpha$ images, SB profiles, intrinsic and scattered Ly$\alpha$ spectra for the same set of simulation runs and snapshots as in \autoref{fig:diff_phys}. In combination, they present a diverse set of Ly$\alpha$ signatures. We see different simulations have completely different intrinsic SB radial profiles, but similar scattered ones. This means the spatially broadening effect of resonant scattering is different in these simulations. Different simulations also have different Ly$\alpha$ spectra giving different values of line parameters. For example, for snapshots with a strong outflow (RTCRiMHD at low $z$), the Ly$\alpha$ spectra have large R2B, low $\sigma_{\rm red}$, whereas snapshots with a weak outflow (RT at high $z$) show the opposite. Significant LOS variations are present for all simulations for both Ly$\alpha$ SB profiles and spectra. We cannot identify systematic trends between Ly$\alpha$ signatures and different physical processes from this small sample of snapshots and LOS so we leave this problem for the statistical study in \autoref{ssec:statistical_studies}.

\begin{figure*}
    \centering
    \includegraphics[width=1.\linewidth]{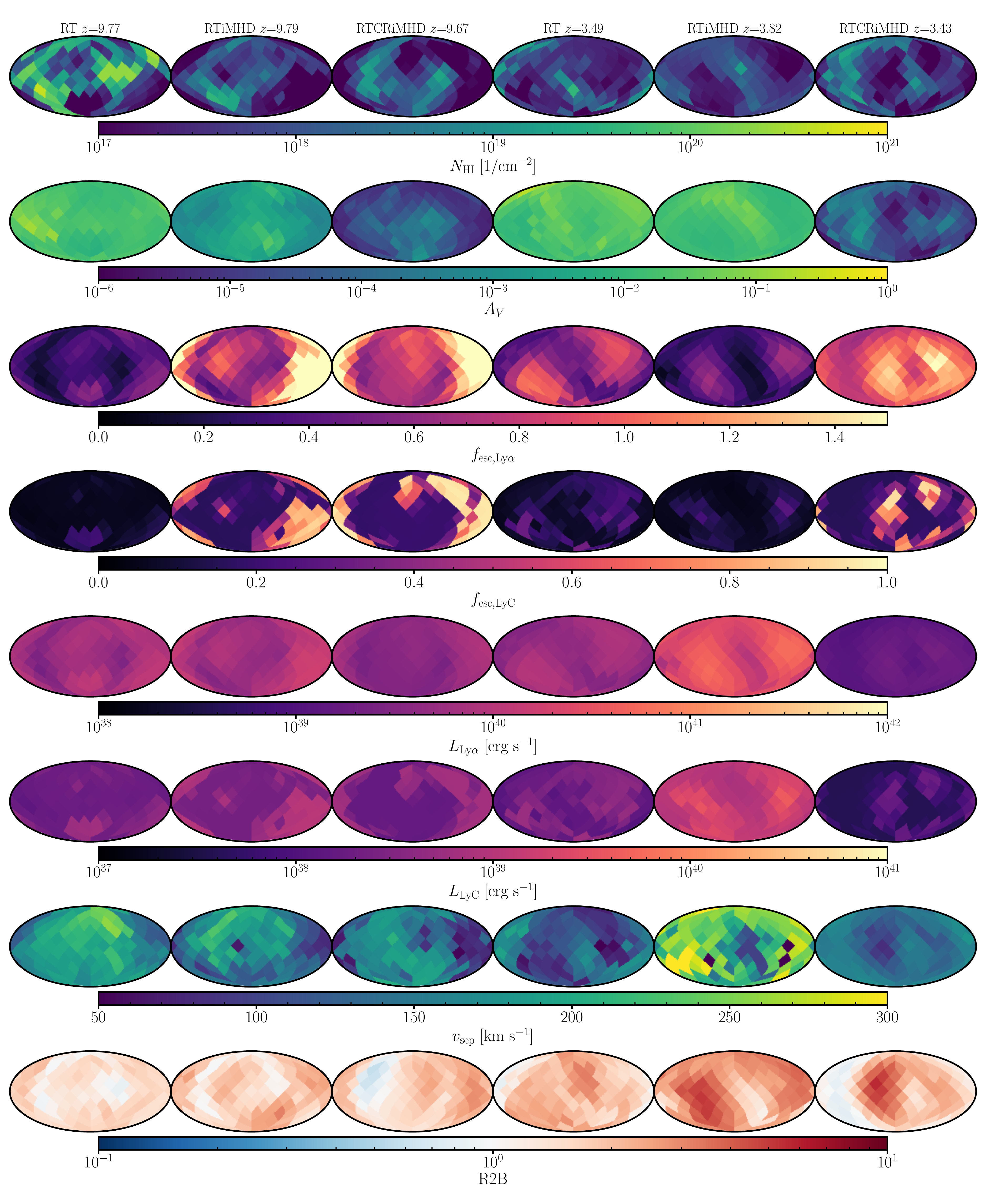}  
    \caption{
    Variations of the angular distribution of various gas properties, Ly$\alpha$ and LyC parameters, due to different simulated physics. From top to bottom panels we show full-sky healpix maps centred on the galaxy for $N_{\rm H~\textsc{i}}$, $A_V$, $f_{\rm esc, Ly\alpha}$, $f_{\rm esc, LyC}$, $L_{\rm Ly\alpha}$, $L_{\rm LyC}$, $v_{\rm sep}$, and R2B. Different columns correspond to different snapshots and simulations, as indicated in the labels in the top panels. The left three columns show the results at `high' $z$ and the right three columns show the results at `low ' $z$. In each redshift bin, the three columns show the results of post-starburst snapshots for the RT, RTiMHD, and RTCRiMHD simulations.}
    \label{fig:ang_dist_diff_phys}
\end{figure*}

We show the full-sky angular distribution of $\NHI$, $A_V$, $\fesclya$, $\fesclyc$, $L_{\rm Ly\alpha}$, $L_{\rm LyC}$ and various Ly$\alpha$ parameters in the form of healpix maps in \autoref{fig:ang_dist_diff_phys}. At `high' redshift, RT has a smooth distribution for all investigated quantities as the mild and consistent feedback is not effective at creating distinct low-density channels. Both RTiMHD and RTCRiMHD have large variations across different LOS as the strong and abrupt feedback creates distinct low-density channels. At `low' redshift, RT and RTiMHD have relatively smooth distributions as they both have entered the self-regulated phase. RTCRiMHD still features disruptive feedback giving large $\fesclya$ and $\fesclyc$. Among all these snapshots, we again see that $b_{\rm LAH}$ and R2B have a similar distribution to that of $\fesclya$ while $\vsep$ correlates with $\fesclyc$. We also see RT and RTiMHD have significantly higher $A_V$ than RTCRiMHD at both `high' and `low' redshift. 

In summary, different physical processes affect Ly$\alpha$ signatures mainly through changing the SFH of the galaxies, the feedback strength, and the gas phase structure. The physical processes affect whether the cold gas can be entrained in the CGM region and the opening angle of low-density channels. Importantly, Ly$\alpha$ spectra are a sensitive probe of the strength of feedback events. With increasing feedback strength, we find Ly$\alpha$ transits from relatively symmetric to red peak dominated profiles.

\subsection{Statistical studies of Ly\texorpdfstring{$\alpha$}{alpha} and LyC escape}
\label{ssec:statistical_studies}

In this section, we review the robustness of the qualitative findings discussed in the last section by exploring a larger sample of LOS and performing further statistical analysis on all the snapshots from all three simulations. We also explore the observational implications of our results. We address two important questions in particular: (1) what are accurate indicators of Ly$\alpha$ and LyC escape? (2) How can we best predict $f_{\rm esc, LyC}$?

\subsubsection{Diagnostics of $\fesclya$ and $\fesclyc$}
\label{sssec:indicators}

The escape of Ly$\alpha$ and LyC photons are both complex processes dependent on ISM properties. It is therefore important to understand this dependence and the connection between Ly$\alpha$ and LyC. 

\begin{figure}
    \centering
    \includegraphics[width=1.\columnwidth]{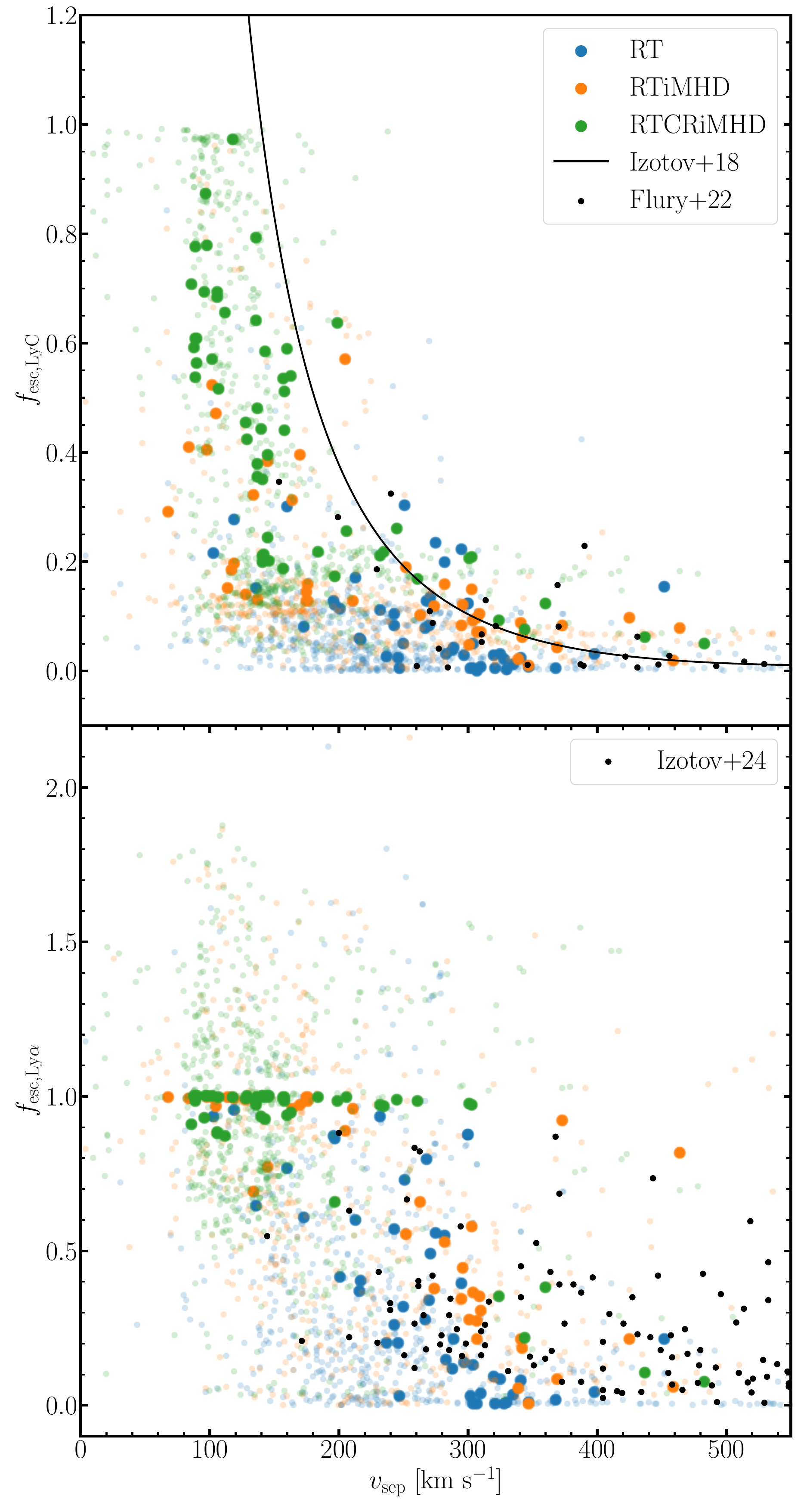}
    \caption{LyC and Ly$\alpha$ escape fractions as a function of $\vsep$ for all the snapshots in the three simulations  with different physics included. Different colours indicate different simulation runs, with a solid point representing the angle-averaged value for a single snapshot. The semi-transparent points show the values along different LOS. We only show 12 different LOS uniformly sampled from the $\fesclyc$ LOS distribution for each snapshot, to display a representative number of data points without cluttering the graph. We further exclude single-peak spectra and only include the spectra with double peaks and triple peaks.  }
    \label{fig:reproduce_Izotov}
\end{figure}

We compare $\fesclyc$ and $\fesclya$ versus $\vsep$ for all snapshots in the three simulations in \autoref{fig:reproduce_Izotov}. The large and solid circles correspond to the angle-averaged values for the studied quantities while the smaller and partially transparent circles represent the values for individual LOS (note that we only show 12 LOS that uniformly sample the $\fesclyc$ LOS distribution for each snapshot to avoid over cluttering the graph). We exclude single-peak spectra to make a better comparison to the observation\footnote{Although in this figure we explicitly exclude the single-peak spectra in our sample, we check that there is actually no such spectrum in the full sample used in this figure.}. 

We first describe the trends observed for angle-averaged values. We start with the correlation between $\vsep$ and $\fesclyc$ \citep{Verhamme15, Yang16, Izotov18, Izotov21, Flury22}. Data points corresponding to different simulation runs occupy different regions in this parameter space. The RT simulation in general has the largest $\vsep$ and lowest $\fesclyc$ whereas the RTCRiMHD simulation displays the opposite trend, with the RTiMHD simulation results populating the intermediate regime. Despite their different distributions, these three simulations all follow  correlations between the angle-averaged $\vsep$ and $\fesclyc$ with different amounts of scatter\footnote{We have performed an experiment where we select the data points based on a luminosity threshold of $10^{39}$~erg\,s$^{-1}$. We found that although the distribution over parameter space is changed, the data still form a universal correlation. This means a luminosity threshold will only change our results quantitatively rather than qualitatively. For simplicity, we do not apply the luminosity threshold in this section.}. This demonstrates that the connection between LyC and Ly$\alpha$ emission is robust to changes in the SF regulating physics. We overplot the fit from \citet{Izotov18} and find that our results reproduce this observed trend reasonably well, although with a small offset. This offset might be due to a number of reasons. Our simulated galaxy has different physical properties than those of the observed sample (e.g. lower masses) and we do not try to mimic the selection effects of the observed sample. We also use a different method of measuring $\fesclyc$. We further note that there is little redshift evolution of the correlations we mention above.

The relation between $\fesclya$ and $\vsep$ displays a much larger scatter than those with $\fesclyc$. $\fesclya$ also saturates at 1, especially for the CR runs. The physical cause is that, although $\fesclyc$ and Ly$\alpha$ line shape parameters are both sensitive to the H~\textsc{i} distribution, $\fesclya$ is mostly sensitive to the amount and distribution of dust, for which our modelling is somewhat simplistic.

The trends discussed above for the angle-averaged values show a much larger scatter when considering the entire LOS distribution displayed (faint points), but the correlations are not completely washed out. We see high $\vsep$ still indicates low $\fesclyc$, but $\fesclyc$ completely spreads out at low $\vsep$, different than in the angle-averaged case. This is due to the orientation effect we have discussed in \autoref{ssec:ori_effect}. Low $\vsep$ values are found for sightlines that partially or completely pierce through low-density channels (category II or III sightlines). However, these two categories can display both low and high values of $\fesclyc$. We further overplot the observational data from \citet{Flury22} and \citet{Izotov24} on the two relations, respectively, and find good agreement with our individual LOS data.

\begin{figure*}
    \centering
    \includegraphics[width=1.\linewidth]{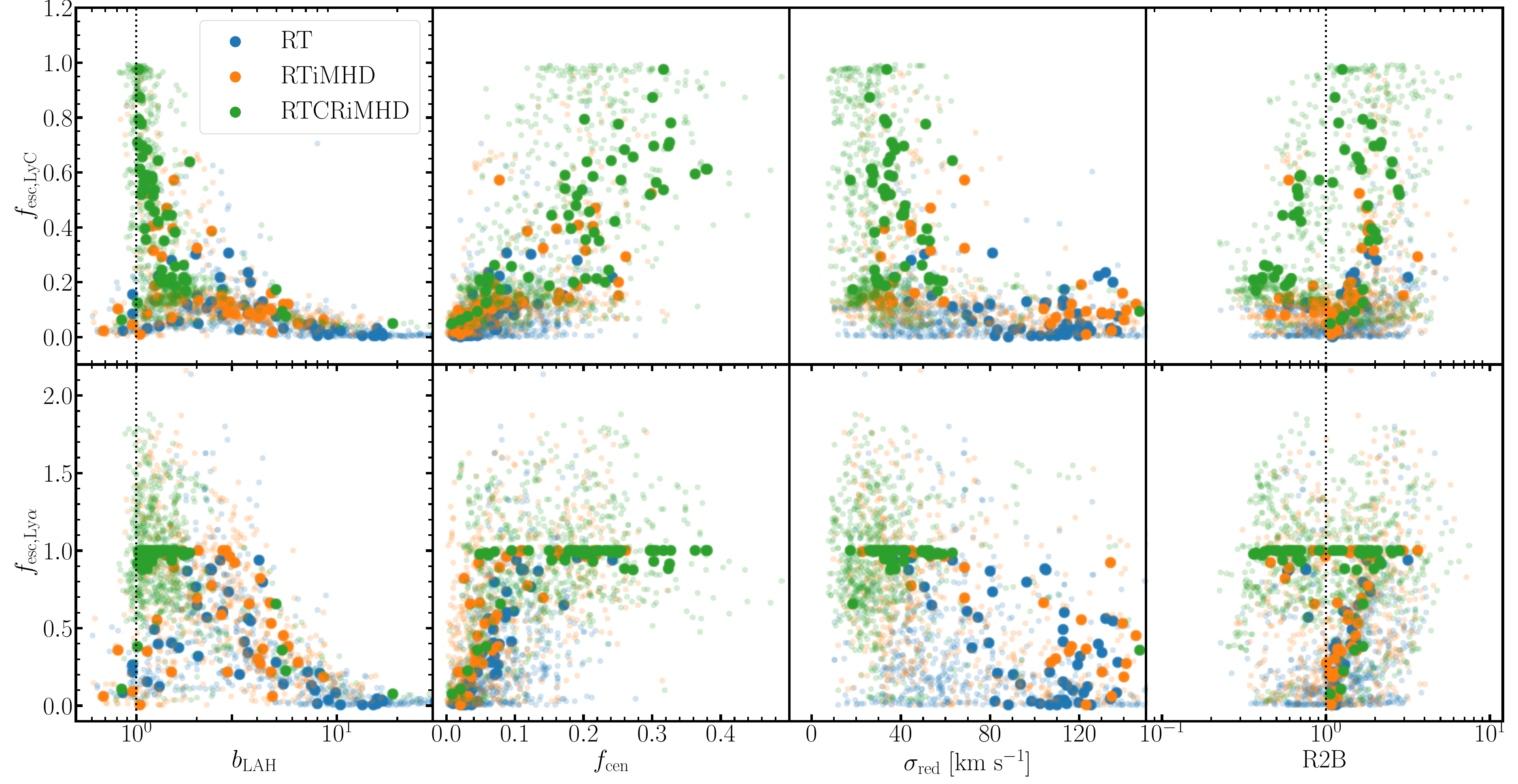}
    \caption{LyC and Ly$\alpha$ escape fractions as a function of the other  Ly$\alpha$ parameters studied. The first row shows $\fesclyc$ versus $b_{\rm LAH}$ (first column), $\fcen$ (second column), $\sigma_{\rm red}$ (third column), R2B (fourth column) for all the snapshots in the three simulations with different physics included. The second row displays the same panels, now for $f_{\rm esc, Ly\alpha}$. The arrangement of this figure is the same as that of \autoref{fig:reproduce_Izotov}. The combined values from the three studied simulations establish universal relations between Ly$\alpha$ parameters and $\fesclyc$, which are somewhat less evident for $\fesclya$. These relations are maintained both for the angle-averaged and individual LOS values, albeit with a larger scatter for the latter.}
    \label{fig:deci_lya}
\end{figure*}

We compare $\fesclyc$ and $\fesclya$ versus various other Ly$\alpha$ parameters for all snapshots in the three simulations in \autoref{fig:deci_lya}. We start with the angle-averaged results.
$\fesclyc$ is in general anti-correlated with the spatial Ly$\alpha$ halo broadening factor $b_{\rm LAH}$. This is because a larger amount of H~\textsc{i} will scatter Ly$\alpha$ photons to larger distances from their emission regions. However, we find that low values of $b_{\rm LAH}$ can correspond to any values of $\fesclyc$ while high values of $b_{\rm LAH}$ require low $\fesclyc$. This means that low $b_{\rm LAH}$ is a necessary but not sufficient condition for high LyC escape. Similarly, we also find the central peak fraction $\fcen$ and the red peak velocity width $\sigma_{\rm red}$ to be relatively accurate indicators of $\fesclyc$. The relation with R2B has significantly more scatter, with a trend for $\fesclyc$ to be generally higher for red peak dominated spectra and less so for blue peak dominated ones. 

The relation between $\fesclya$ and those other Ly$\alpha$ parameters displays a much larger scatter than those with $\fesclyc$. However, we see a clear trend that high $b_{\rm LAH}$ indicates low $\fesclya$. Another trend we find is that R2B sets a lower limit for $\fesclya$ at both ends. The physical reason behind this is somewhat complex: high R2B values correspond to post-starburst phases with high $\fesclya$ due to the development of low density channels, while low R2B values correspond to inflow phases with little dust content and high $\fesclya$.

Similar to the case of $\vsep$, we see a much larger scatter for the full LOS samples. Nonetheless, much of our individual LOS data follow the same correlations as the angle-averaged measurements. In particular, we see two trends remain robust: (1) high $b_{\rm LAH}$ and high $\sigma_{\rm red}$ both indicate low $\fesclyc$; (2) high $b_{\rm LAH}$ indicates low $\fesclya$. 

\begin{figure}
    \centering
    \includegraphics[width=1.\columnwidth]{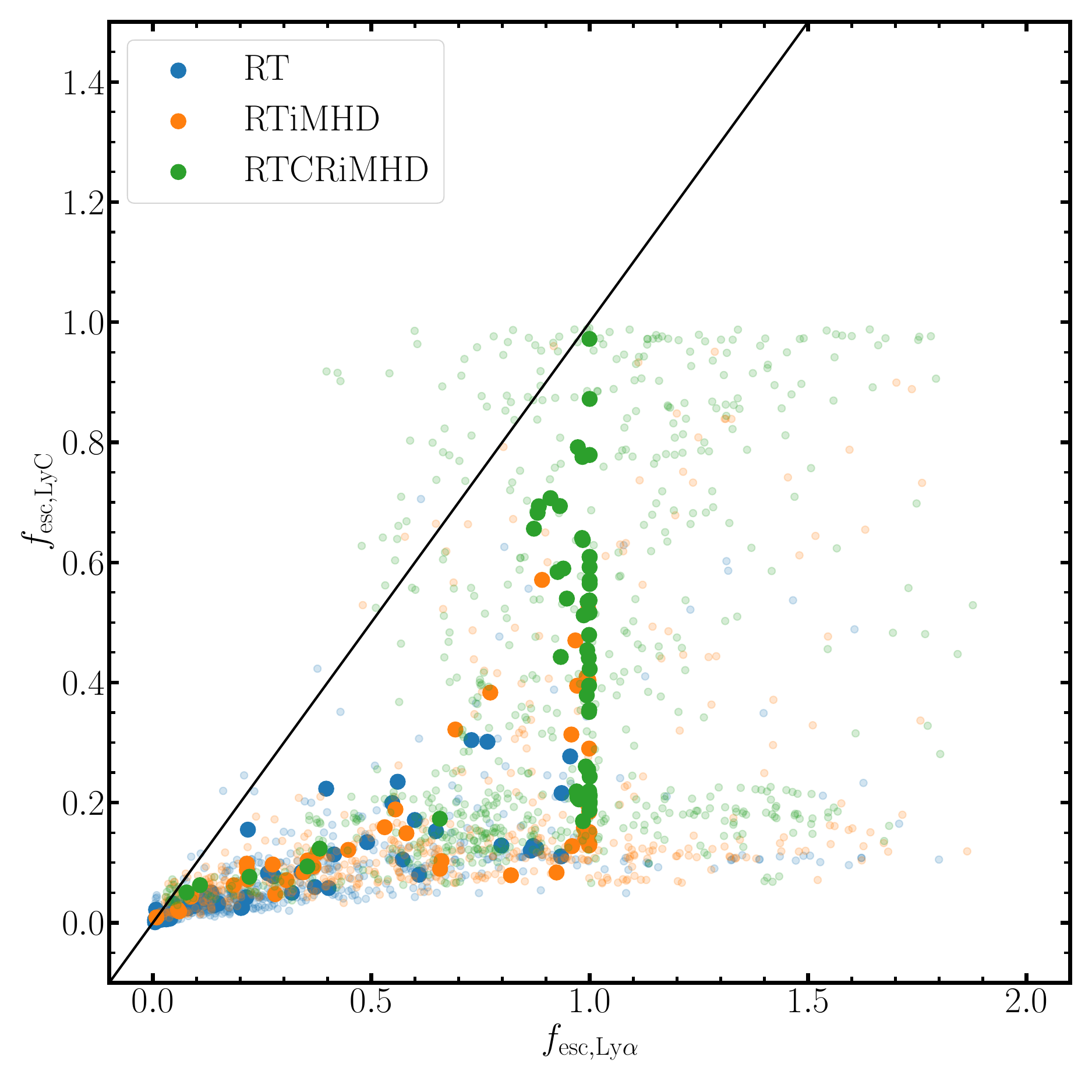}
    \caption{$\fesclyc$ versus $\fesclya$ for all snapshots in the three simulation runs. The arrangement of this figure is the same as that of \autoref{fig:reproduce_Izotov}, with the solid black line displaying the 1:1 relation.}
    \label{fig:lya_lyc}
\end{figure}

We next compare Ly$\alpha$ versus LyC escape in \autoref{fig:lya_lyc}. For both angle-averaged and individual LOS escape fractions, the values of $\fesclyc$ are always smaller than those of $\fesclya$ (very few exceptions in the individual LOS data). In general, $f_{\rm esc, LyC}$ increases with $f_{\rm esc, Ly\alpha}$ at $\fesclya < 75\%$. As $f_{\rm esc, Ly\alpha}$ saturates at unity, $\fesclyc$ spreads out between $\sim 0.1 \to 1$ (cf. Figure 17 in \citealt{Katz22c}). 

Distinguishing different models observationally, can, at least in principle, be done in the hyper-parameter space spanned by gas, Ly$\alpha$ and LyC parameters using a machine learning classification algorithm. However, this will require (i) a much larger simulation suite with multiple variants of physical models and parameters; and (ii) careful descriptions of telescope effects. Both of these requirements are beyond the scope of this paper so we left this for future work. 

\subsubsection{Ly\texorpdfstring{$\alpha$}{alpha} based prediction of $\fesclyc$ }

In this section, we aim to find a quantitative method to predict $\fesclyc$ based on all the Ly$\alpha$ parameters we investigated in \autoref{sssec:indicators}, for the stellar mass range of our simulated dwarf galaxies. We express LyC escape as a function of various line parameters 

\begin{equation}
\fesclyc = f( b_{\rm LAH}, \vsep, \fcen, \sigma_{\rm red}, \mathrm{R2B}, \fesclya)\,.
\end{equation}

\begin{figure}
    \centering
    \includegraphics[width=1.\columnwidth]{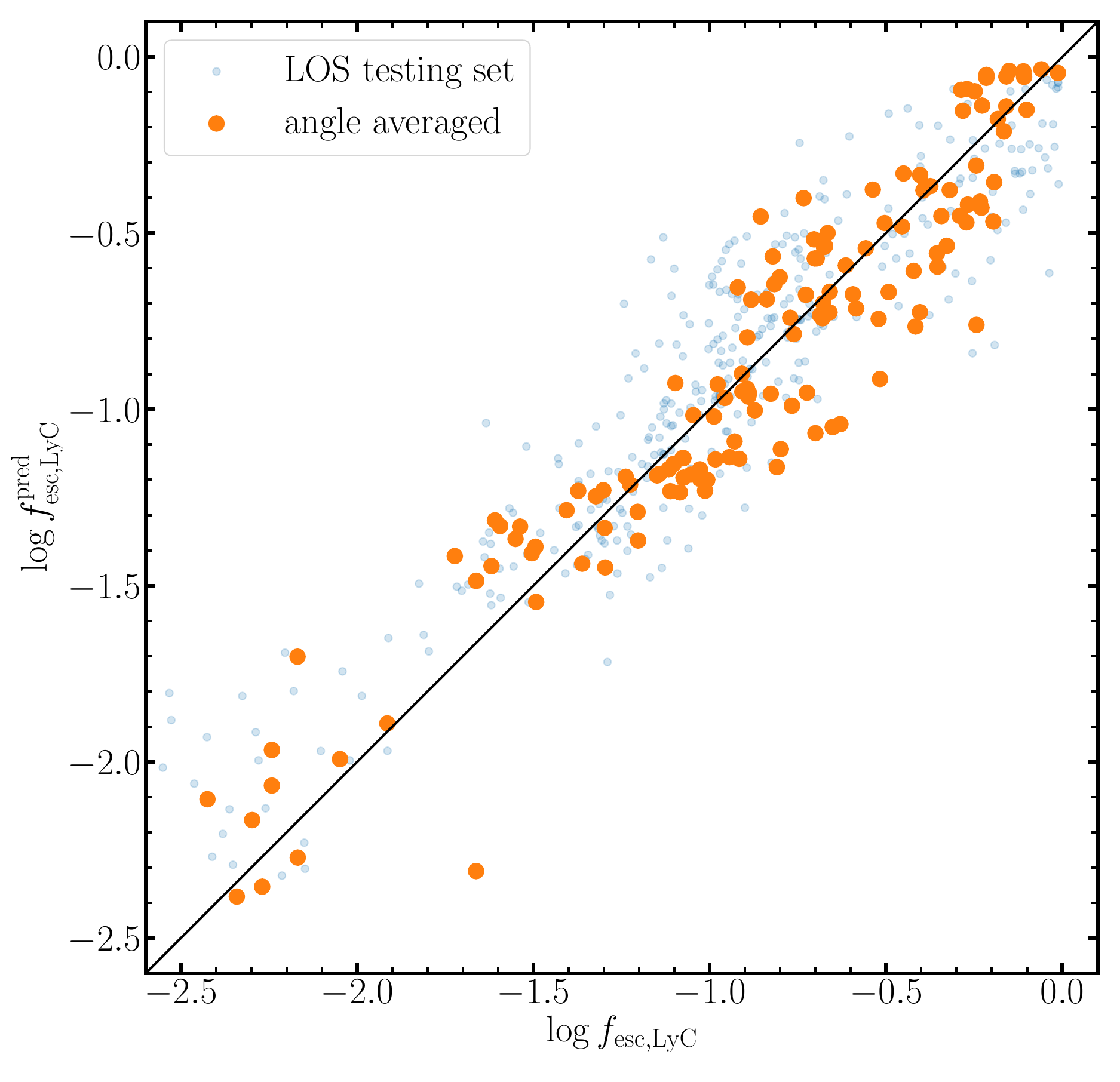}
    \caption{$\fesclyc$ predicted from Ly$\alpha$ observables by our random forest regression formula versus actual $\fesclyc$ values, for all snapshots in the three simulation runs. The semi-transparent blue points correspond to the training set of the individual LOS data, while solid orange points correspond to the angle-averaged data. The black line shows the 1:1 relation.}
    \label{fig:pred_fesc}
\end{figure}

We have tried multiple regression algorithms (linear model, decision trees, and random forest), and find that the random forest achieves the best results among these algorithms. We therefore use the random forest algorithm implemented in \textsc{SCIKIT-LEARN} to fit a single regression function to our data jointly for all three simulations and for all individual LOS. We do not set the maximum depth of the tree and normalize all the parameters. We employ 80\% of our data as a training set and reserve the remaining 20\% as a testing set. Our resulting fit has a rms error of about $\sim 0.02$ for the predicted $\fesclyc$, which means it gives a good representation of the training data. We next investigate the quality of our fit using the testing set and explore whether our fitted function from the individual LOS data applies to angle-averaged data. In \autoref{fig:pred_fesc} we show the predicted $\fesclyc$ versus actual $\fesclyc$ values. Blue points show the training set of individual LOS data, while orange points show the angle-averaged data. We find the regression function to provide reasonably accurate predictions for both the individual LOS and angle-averaged data. This demonstrates that (i) the connection between Ly$\alpha$ parameters and $\fesclyc$ is robust against varying the included simulated physics; (ii) for our simulations we can use the properties we have chosen as a proxy for the individual LOS LyC escape to infer the angle-averaged LyC escape if the individual LOS sample size is large enough.
We also rank the importance of the parameters returned by random forest regression (see \citealt{Bluck20a, Bluck20b, Piotrowska22} for details). The relative importances of all six parameters are (ranked from important to unimportant): ($\bLAH$, 0.32), ($\fesclya$, 0.20), ($\fcen$, 0.20), ($\vsep$, 0.18), ($\sred$, 0.05), (R2B, 0.05). We see that $\bLAH$ is the most important one, with $\fesclya$, $\fcen$ and $\vsep$ as secondary predictive parameters. The remaining parameters $\sred$ and R2B are of low importance.

\section{Discussion}
\label{sec:discussion}

\subsection{Ly\texorpdfstring{$\alpha$}{alpha} duty cycle}

Having analyzed the physical origin of the diversity of Ly$\alpha$ signatures, we are now in a position to investigate how the duty cycle of our dwarf galaxies is traced by Ly$\alpha$ and LyC emission. \citet{Naidu22} proposed a unified model combining models from \citet{Tenorio-Tagle99, Mas-Hesse03, Mao07} and several simulations. We give a brief description of their model here.  \citet{Naidu22} distinguish four phases during the evolution of high-z galaxies. In phase I, a symmetric double peaked Ly$\alpha$ profile is predicted as an initial starburst occurs but the subsequent feedback has not yet accumulated enough energy to disrupt dense clouds. In phase II, triple-peaked profiles emerge, as feedback destroys the surrounding clouds, creating low density channels and giving rise to high Ly$\alpha$ and LyC escape\footnote{We note that the triple peak profile mentioned in \citet{Naidu22} refers to a superposition of a double peak profile resulting from resonant scattering and a single peak centred near zero velocity from the direct escape of Ly$\alpha$ photons via low-density channels. This is different from the triple peak profile we measure from our simulations that is due to complex motion of gas or deuterium, and we never see a single peak centred near zero velocity.}. In phase III, the spectra show a prominent red peak as the dense structure starts to reform in the presence of much milder star formation. Finally, in phase IV, the spectra are much fainter and smoothed out to the line wings, as star formation gets quenched and dust and gas are assembled in the surroundings of galaxies. The observations presented in \citet{Naidu22} suggest that 50\% of LAEs are in phase II, either because galaxies are observed along specific sightlines or because consecutive starbursts shorten the duration of phase III and repeatedly shift the galaxy cycle back to phase II. However, it remains unclear whether selection bias effects or physical effects have a larger impact on determining the fraction of observed LAEs in different phases. 

Our results qualitatively agree with the picture outlined by \citet{Naidu22}. However, the picture from our simulations is further complicated by line-of-sight effects, galaxy mergers, the characteristic variability of dwarf galaxy evolution due to their shallow gravitational potential, and how the gas distribution is affected by the different physical processes. Mergers and SF histories affect the duration of the different phases. Different gas morphology and covering fractions created by different feedback mechanisms will affect the fraction of sightlines that can be observed, $f_{\rm LOS}$. Motivated by the above statements we analyse two limiting cases from our simulations. In the RT simulation, smooth, self-regulated SFH makes the time duration of each phase comparable to each other. Intrinsic luminosity and sightline fraction $f_{\rm LOS}$ do not evolve strongly and the dwarf galaxy can be visible in all phases. In the RTCRiMHD simulation, the bursty merger-driven SFH makes phase II significantly shorter than the other phases. During phase II, both the intrinsic luminosity and sightline fraction $f_{\rm LOS}$ are significantly higher, and the galaxy 
is only visible during this phase.

\subsection{Caveats}

Although our modelling adopts arguably the most complete set of physical processes while reaching the highest resolution to date for such simulations, several caveats still remain that could affect the interpretation of our results.

We first note that the spatial and mass resolution of the \textit{PANDORA} simulation suite still cannot fully resolve the multi-phase ISM and the ionizing photon escape, which will affect the thermal state of the ISM and the simulated line emission. Even higher spatial resolution than the employed $\Delta x \sim 7$~pc is required to fully resolve individual low-density channels, a turbulent porous ISM and individual GMC and HII regions \citep{JG_Kim19, Kimm19, Kakiichi&Gronke21, Martin-Alvarez22}. Our refinement scheme is configured to resolve both the quasi-Lagrangian mass evolution and the Jeans length of gas parcels with at least 4 cells. Consequently, it does not resolve all the potential Strömgren sphere in the ISM \citep{JG_Kim19, Smith21b} and small-scale, multi-phase structures that may exist in the CGM \citep{McCourt18, Bennett20}. Our chemistry and radiative heating and cooling modelling are also quite simplistic, future work should incorporate time-dependent, non-equilibrium chemistry coupled with radiative transfer \citep{Katz22b, Kim_JG23}. Our star formation and stellar population model only samples a full stellar particle more massive than individual stars which may reduce the level of inhomogeneous and stochastic stellar feedback for dwarf galaxies \citep{Smith21b}. SFR stochasticity also induces a large scatter to Ly$\alpha$ and LyC production \citep{Forero-Romero13}. All the above effects will affect the stochasticity of our results and the LOS statistics. Furthermore, we still have little knowledge of the microphysics of CRs \citep{Bai22}. The next generation of galaxy formation models will hopefully give us a better understanding of this. The dust modelling is also rather simplified in our simulations as we only assume that dust passively traces metals in the warm neutral gas, which could affect how Ly$\alpha$ photons are absorbed within the galaxy. 

The next set of caveats regards the simulation volume and statistics. As we use a zoom-in simulation of a single dwarf galaxy, we only probe the low-mass-end of the high-z galaxy population and neglect the effect of variety in the environment \citep{Byrohl21} as well as the effect of IGM transmission \citep{Weinberger18, Aaron_Smith22, Garel21}.  Higher mass galaxies will have different density distributions, morphologies, and environments than the dwarf galaxies we study here, which affects how Ly$\alpha$ is generated and transported. Whether our conclusions will hold for  larger galaxy masses is uncertain and will have to be studied in future work. Ly$\alpha$ emission from sources before reionization ends (at $z\approx 6$) will suffer strong IGM attenuation on its way to the telescope (see \citet{Keating24} for a recent discussion). However, a surrounding ionized bubble may allow  Ly$\alpha$ redshifted out of resonance to travel through the IGM. To self-consistently model this effect, however, requires full cosmological RT simulations that can capture a range of galaxy populations and reionization histories of the Universe. Our zoom-in simulations, on the other hand, can only capture local ionization effects  and neglect external ionization by possibly more massive galaxies. 

We last note that simulating radiative processes and generating realistic mock observations is highly non-trivial. All observational facilities have a finite aperture that will smooth out the image with a point spread function (PSF), and due to their finite sensitivity and resolution observations will often not detect the faint emission in the line wings of the spectra and in the outskirts of the galaxies. 

\section{Conclusions}
\label{sec:conclusions}

In this paper, we systematically investigate the use of Ly$\alpha$ as a sensitive tracer of LAE evolution and the physical processes regulating Ly$\alpha$ emission and LyC escape. For this purpose, we employ the \textit{PANDORA} simulation suite, which contains one of the currently most complete set of simulations of a dwarf galaxy with different combinations of physical processes. We post-process our simulations with the parallel radiative transfer code \textsc{RASCAS} and compare synthetic observations with the properties of observed Ly$\alpha$ emitters. Our main findings are listed below. 

\begin{itemize}
    \item Ly$\alpha$ haloes in our simulations generally show a spatially extended structure due to the resonant scattering by the widespread cold neutral medium. This will introduce spatial offsets between Ly$\alpha$ and nebular line haloes. The Ly$\alpha$ spectra show a diverse set of profiles that are sensitive to the dynamical state of the cold neutral gas in the galaxy. A relatively static gas clump will give a double-peaked profile, while outflows are always associated with red peaks. 
    
    \item Ly$\alpha$ emission undergoes anisotropic scattering and varies strongly along different LOS. The level of variation depends on the inhomogeneous distribution of H~\textsc{i} and dust that is strongly affected by different feedback processes. The escape channels of LyC emission are narrower than those of the Ly$\alpha$ emission. This motivates us to distinguish three categories of sightlines: category I LOS are completely obscured by foreground H~\textsc{i} gas and have low $\fesclyc$ and low $\fesclya$; category II LOS are only partially obscured, and have low $\fesclyc$ and high $\fesclya$; category III LOS fully pierce through low-density channels and therefore have high $\fesclyc$ and high $\fesclya$. 

    \item Varying the simulated physical processes has a large impact on Ly$\alpha$ signatures and the LyC escape of dwarf galaxies. Our RT simulation features self-regulated SFH and the weakest feedback, hence the spectra show a relatively symmetric profile with larger values of $\vsep$.  Our RTCRiMHD simulation, on the contrary, shows a bursty SFH and the most prominent cold outflows, resulting in the lowest $\vsep$ and the largest $\fesclyc$ across our simulation suite. The RTiMHD simulation sits in between RT and RTCRiMHD.

    \item Universal correlations between $\fesclyc$ and various Ly$\alpha$ line shape parameters are found across all the snapshots from all three simulation runs, both for the angle-averaged results and the individual LOS results, albeit with a large scatter for the latter. We recover in particular the anti-correlation between $\vsep$ and $\fesclyc$ found in \citet{Izotov18}. We furthermore find a random forest regression function relating $\fesclyc$ to a variety of  Ly$\alpha$ line shape parameters for individual LOS data. Interestingly, this regression function also applies to angle-averaged data, suggesting that the properties of angle-averaged $\fesclyc$ can be obtained from sufficiently large samples of individual LOS data.
    
\end{itemize}

Overall, we conclude that the escape of Ly$\alpha$ emission from dwarf galaxies is extremely sensitive to the interplay of the often merger-driven and generally very bursty star formation history of dwarf galaxies with their ISM and  CGM. Despite the complex dependence of the spatial and spectral distribution of the Ly$\alpha$ emission on the spatial distribution and the dynamical state of neutral hydrogen, Ly$\alpha$ observables of sufficiently large samples of dwarf galaxies will thus provide important constraints on the physical mechanism responsible for galactic outflows as well as the stellar feedback-regulated escape of the LyC emission from what are likely to be the main drivers of the reionization of hydrogen.

\section*{Acknowledgements}

YY thanks for discussions with Joris Witstok, Charlotte Simmonds, Harley Katz and Joki Rosdahl. YY also thanks for the comments on this manuscript from Harley Katz. YY is supported by an Issac Newton's studentship from the Cambridge Trust. SMA acknowledges support by a Kavli Institute for Particle Astrophysics and Cosmology (KIPAC) Fellowship. 

The Pandora simulation suite was supported by ERC Starting Grant
638707 “Black holes and their host galaxies: co-evolution across
cosmic time”. SMA, DS and MGH acknowledge support from
the UKRI Science and Technology Facilities Council (grant number
ST/N000927/1). TG is supported by ERC Starting grant
757258 ‘TRIPLE’.

Both the Pandora simulation suite and the RASCAS radiative transfer simulations in this work were performed on the DiRAC@Durham facility managed by the Institute for Computational Cosmology on behalf of the STFC DiRAC HPC Facility (www.dirac.ac.uk). The equipment was funded by BEIS capital funding via STFC capital grants ST/P002293/1, ST/R002371/1 and ST/S002502/1, Durham University and STFC operations grant ST/R000832/1. DiRAC is part of the National e-Infrastructure. This work was performed using resources provided by the Cambridge Service for Data Driven Discovery (CSD3) operated by the University of Cambridge Research Computing Service (www.csd3.cam.ac.uk), provided by Dell EMC and Intel using Tier-2 funding from the Engineering and Physical Sciences Research Council (capital grant EP/P020259/1), and DiRAC funding from the Science and Technology Facilities Council (www.dirac.ac.uk).

\section*{Data availability}
The data underlying this article will be shared on reasonable request to the corresponding author.




\bibliographystyle{mnras}
\bibliography{refs}



\appendix

\section{Convergence test}
\label{sec:conv_test}

\begin{figure*}
    \centering
    \includegraphics[width=1.\linewidth]{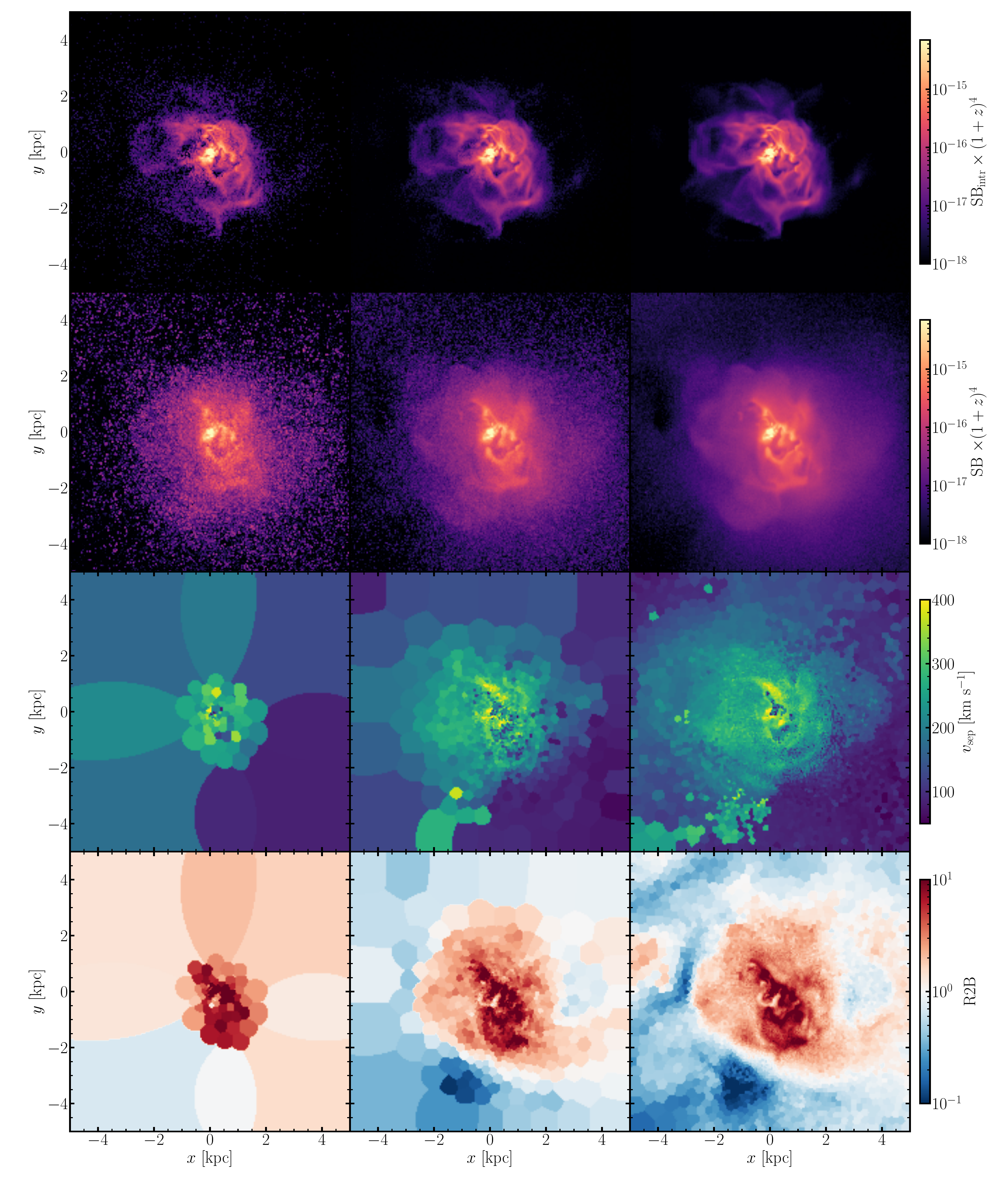}
    \caption{Convergence test of photon counts for 4 types of Ly$\alpha$ images. The three columns show the results for $N_{\rm ph} = 2\times 10^4$ (left), $2\times 10^5$ (default; middle), and $2\times 10^6$ (right). The four rows show intrinsic SB maps (first), scattered SB maps (second), spatially resolved $\vsep$ (third) and R2B maps (fourth).}
    \label{fig:img_conv_test}
\end{figure*}

\begin{figure}
    \centering
    \includegraphics[width=1.\columnwidth]{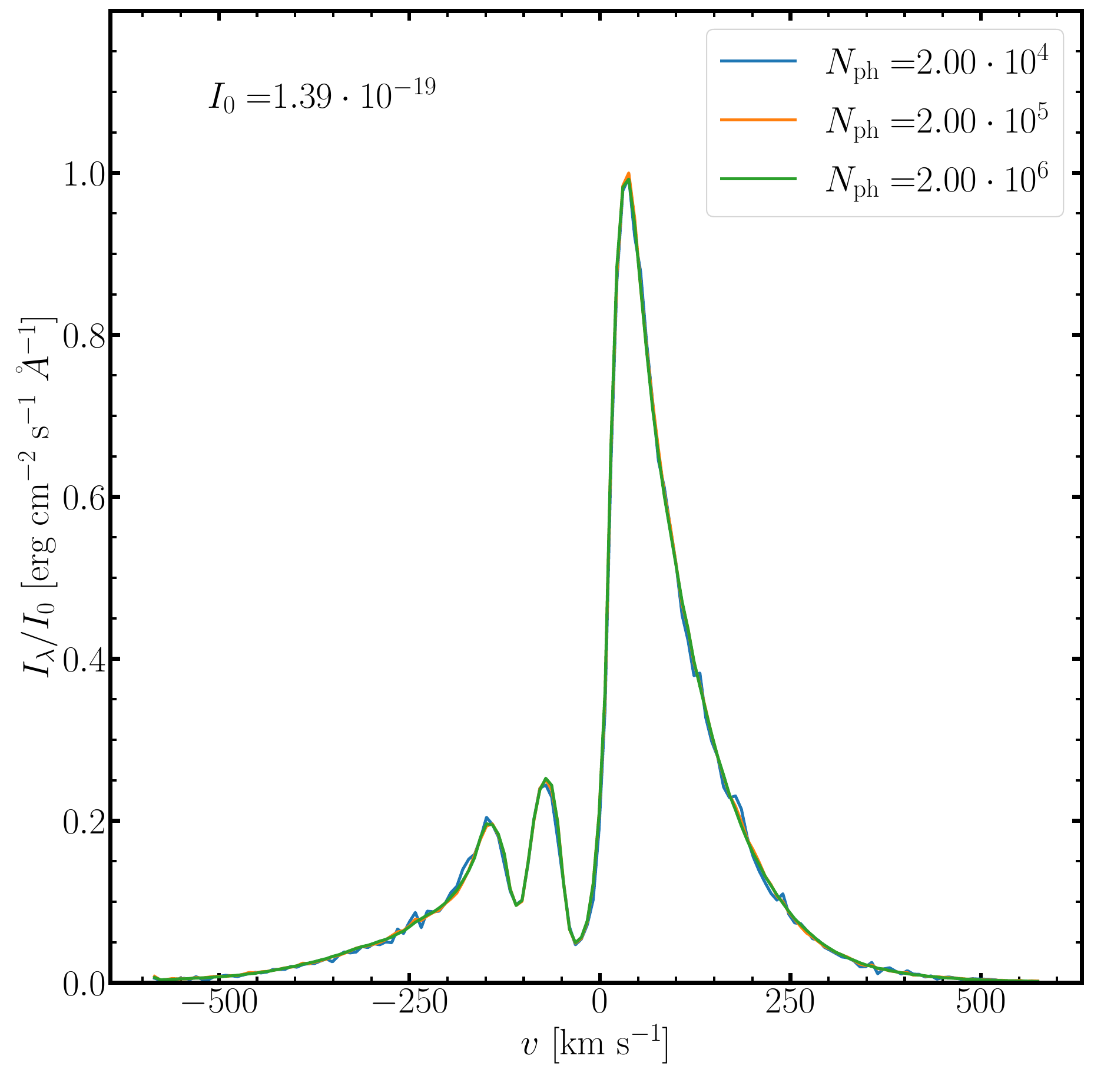}
    \caption{Convergence test of photon counts for Ly$\alpha$ spectra. The three lines show the results for $N_{\rm ph} = 2\times 10^4$, $2\times 10^5$, and $2\times 10^6$.}
    \label{fig:spec_conv_test}
\end{figure} 

We have performed convergence tests for our \textsc{RASCAS} simulations. In \autoref{fig:img_conv_test} we show the intrinsic and scattered Ly$\alpha$ images and spatially resolved $\vsep$ and R2B maps along one LOS for RTCRiMHD simulation at $z=3.49$, for photon packet numbers of $2 \times 10^4$, $2\times 10^5$ (default), $2\times 10^6$, while in \autoref{fig:spec_conv_test} we show the corresponding results for Ly$\alpha$ spectra. Compared with the highest resolution simulation, we see that at the fiducial photon counts, the Ly$\alpha$ images, spatially resolved parameter maps, and spectra all capture the essential observable structures. Further increasing photon counts will only better resolve the outskirts and reducing photon counts will lose important information at the galactic centre. We therefore can say that our photon number counts give a reasonable balance between numerical accuracy and computational efficiency.

\section{$\fesclya$ and $\fesclyc$ of other physical models}
\label{sec:other_model}

\begin{table*}
  \centering
  \begin{tabular}{ccccc}
\hline
Simulation name & $\langle \fesclya \rangle_t$ & $\langle \fesclya \rangle_{L_{\rm Ly\alpha} }$ & $\langle \fesclyc \rangle_t$ & $\langle \fesclyc \rangle_{\rm L_{\rm LyC} }$ \\ \hline
RT & 0.32 & 0.35 & 0.07 & 0.16 \\
RT+Boost & 0.11 & 0.17 & 0.00 & 0.02 \\
RTiMHD & 0.61 & 0.43 & 0.16 & 0.16 \\
RTsMHD & 0.49 & 0.36 & 0.10 & 0.13 \\
RTnsCRiMHD & 0.81 & 0.47 & 0.34 & 0.19 \\
RTCRiMHD & 0.89 & 0.26 & 0.47 & 0.20 \\
\hline
\end{tabular}
  \caption{Time-averaged and Ly$\alpha$ luminosity-weighted mean $\fesclya$, as well as time-averaged and LyC luminosity-weighted mean $\fesclyc$, for all six simulation runs.}
  \label{tab:fesc_other_model}
\end{table*}

In the main body of the paper, we have seen that the different simulated physics results in  drastically different values and histories of $\fesclya$ and $\fesclyc$. To explore this further we additionally run \textsc{RASCAS} simulations for Ly$\alpha$ and LyC for the remaining \textit{PANDORA} simulations RT+Boost, RTsMHD, and RTnsCRiMHD. For simplicity, we only focus on the angle-averaged results. 

We show time-averaged and Ly$\alpha$ luminosity-weighted mean $\fesclya$, as well as time-averaged and LyC luminosity-weighted mean $\fesclyc$ in \autoref{tab:fesc_other_model}. Once again, we see completely different levels of Ly$\alpha$ and LyC escape for different simulated physics, with RT+Boost the lowest and RTCRiMHD the highest. The time-weighted mean values are generally different from luminosity-weighted ones and this discrepancy highly depends on the SFH and feedback-regulated duty cycles of the galaxies. 

\end{document}